\documentclass[12pt]{article}
\usepackage{styles/mystyle}

\begin{document}
    \renewcommand*{\thefootnote}{\fnsymbol{footnote}}

    \title{TechRank}\footnotetext{This document is the results of a research project funded by the Cyber-Defence Campus, armasuisse Science and Technology.}
    \date{}

    \renewcommand*{\thefootnote}{\arabic{footnote}}

    \author[1]{Anita {Mezzetti}}
    \author[2,3]{Loïc Maréchal}
        
    \author[4]{Dimitri {Percia David}}
        
    \author[2]{William Lacube}
     
    \author[5,6]{Sébastien Gillard}
            
    \author[2]{Michael Tsesmelis}
    
    \author[5]{Thomas Maillart}
        
    \author[2]{Alain Mermoud}

    \affil[1]{\small Credit Suisse}
    
    \affil[2]{Cyber-Defence Campus, armasuisse Science and Technology}
    
    \affil[3]{University of Lausanne (HEC Lausanne)}
    
    \affil[4]{University of Applied Science (HES-SO Valais-Wallis)}

    \affil[5]{GSEM, University of Geneva}

    \affil[6]{Military Academy, ETH Zurich}
    
\singlespacing
\maketitle
\vspace{-.2in}

    \begin{abstract}
       We introduce TechRank, a recursive algorithm based on a bi-partite graph with weighted nodes. We develop TechRank to link companies and technologies based on the method of reflection. We allow the algorithm to incorporate exogenous variables that reflect an investor's preferences. We calibrate the algorithm in the cybersecurity sector. First, our results help estimate each entity's influence and explain companies' and technologies' ranking. Second, they provide investors with a quantitative optimal ranking of technologies and thus, help them design their optimal portfolio. We propose this method as an alternative to traditional portfolio management and, in the case of private equity investments, as a new way to price assets for which cash flows are not observable.
    \end{abstract}

\medskip


\thispagestyle{empty}
\clearpage


\section{Introduction}\label{sec1}
This work investigates the innovation structure and the dynamics underlying the life cycle of technologies. We fill two research gaps. The first concerns the identification of future benefits and risks of emerging technologies for the society. The second regards the valuation or early-stage companies and optimal investment decisions. To fill these gaps, we introduce the TechRank algorithm. Our methodology assigns a score to each entities, \textit{i.e.}, technologies and firms, based upon their contribution to the technological ecosystem. We expect this method to help stakeholders in forming optimal decisions for investment, procurement, and technology monitoring.

We calibrate our model in the cybersecurity sector, although TechRank could apply to any sector. The cybersecurity technological landscape represents a particular challenge for this calibration, given the important share of start-ups and fast innovations it yields.\cite{GordonLoebLucyshynZhou2018}. Moreover, the important number of cyber-attacks and the increasing costs they incur has boosted cybersecurity investments.\footnote{
\href{https://www.nytimes.com/2021/07/26/technology/cyberattacks-security-investors.html}{The New York Times: \quotes{As Cyberattacks Surge, Security Start-Ups Reap the Rewards}} by Erin Woo (July 26, 2021).\\
\href{https://finance.yahoo.com/news/microsoft-securing-position-cybersecurity-investments-082833371.html}{Yahoo Finance: \quotes{Microsoft Securing its Position with Cybersecurity Investments}} by TipRanks (July 20, 2021).} According to Bloomberg, \quotes{the global cybersecurity market size is expected to reach USD 326.4 billion by 2027, registering a compound annual growth rate of 10.0\% from 2020 to 2027}.\footnote{\href{https://www.bloomberg.com/press-releases/2020-07-29/global-cybersecurity-market-could-exceed-320-billion-in-revenues-by-2027}{Bloomberg: \quotes{Global Cybersecurity Market Could Exceed \$320 Billion in Revenues by 2027}} (July 29, 2020).}

To develop the TechRank algorithm, we first model and map the ecosystem of companies and technologies from the Crunchbase dataset using a bi-partite network. The bi-partite network structure accurately describes this complex and heterogeneous system. We evaluate the relative influence of the network nodes in the ecosystem by adapting a recursive algorithm that estimates network-centrality.

This methodology should help decision-makers and investors to assess the influence of entities in the cybersecurity ecosystem, reducing investment uncertainties. In fact, around 90\% of startups fail and in 42\% of the cases this is due to incorrect evaluation of the market demand. The second reason (29\%) is because they run out of funding and personal money.\footnote{\href{https://findstack.com/startup-statistics/}{Findstack: \quotes{The Ultimate List of Startup Statistics for 2021}} by Jack Steward.} Christensen (1997) highlights that well-managed companies also break down, because they over-invest in new technologies\cite{Christensen1997}. Thus, by selecting the right technologies to invest in goes along with the optimal investment strategy.

Our research takes inspiration from Google's PageRank algorithm, that ranks web pages according to readers' interest\cite{PageBrinMotwaniWinograd1999}. We use a similar approach with bi-partite networks to assign a score to companies and technologies. Our method is flexible and permits to incorporate investors' preferences such as location or previous funding rounds. TechRank let the investor select entities' features that reflect their interests. The algorithm uses their choices as input, which tweaks the entities' score to reflect them. This enables investors to select a personalized portfolio strategy using a quantitative methodology. The evaluation of companies and new technologies largely depends on investors' personal choices, which may lead to misread market demand. This work aims to lead to more methodical decision making for investors.

The remainder of this article proceeds as follows. Section \ref{sec2} presents the literature review and hypotheses. Section \ref{sec3} details the data and the methodology. Section \ref{sec4} presents the results. Section \ref{sec5} concludes.
\section{Literature review and hypotheses development}\label{sec2}
\subsection{Centrality measures}

In network analysis, the centrality estimates the importance of nodes through ranking. The most simple centrality estimate is the \quotes{degree}, which counts the number of neighbours of a node. One of its drawbacks is that it does not show which one stands in the center of the network. Two nodes may share the same degree, while being more or less peripheral. Thus, the degree is a local centrality measure, which does not capture the influence across nodes within the graph.

\begin{figure}[h!]
    \centering
    \includegraphics[scale=0.4]{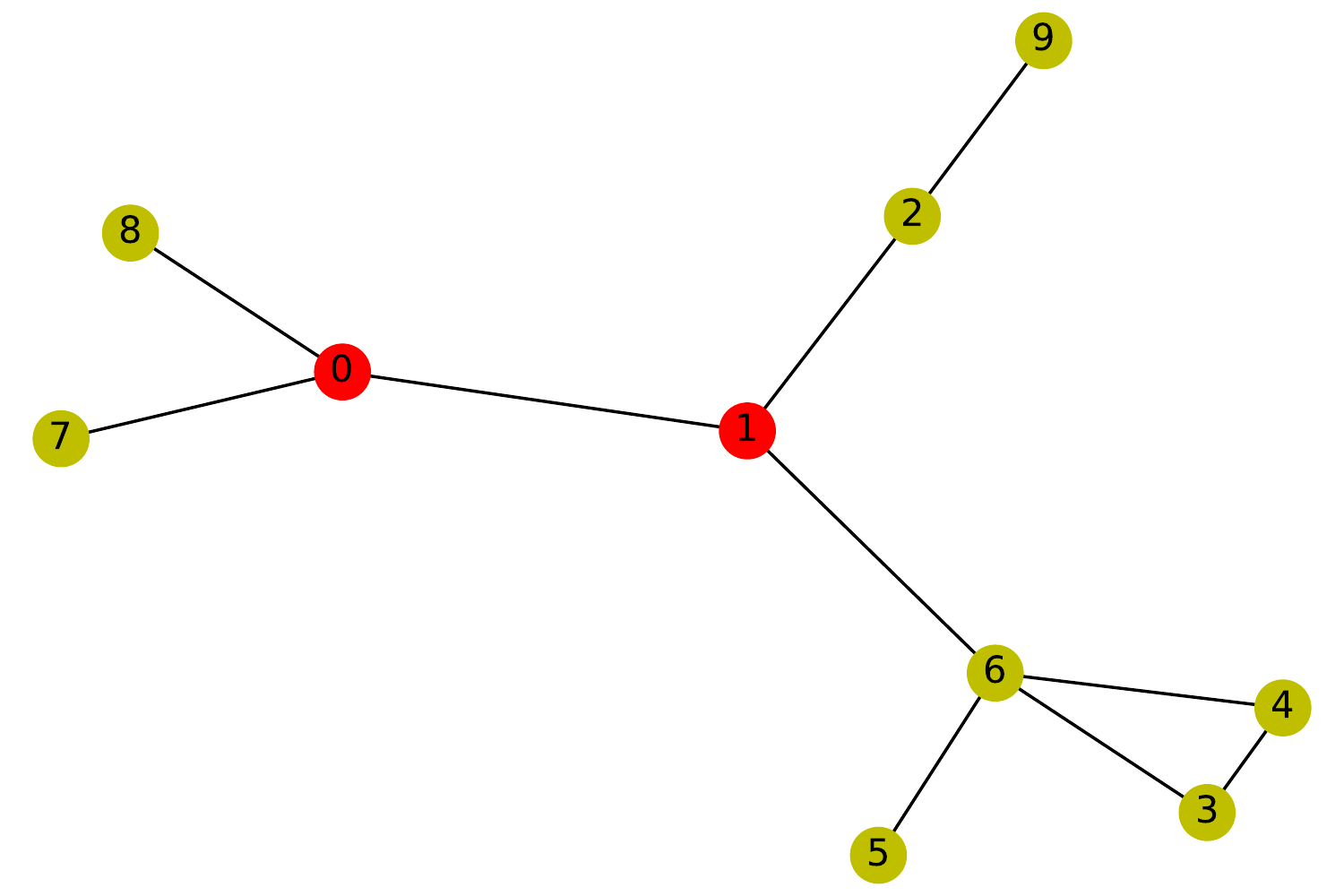}
    \caption{Central and peripheral nodes}
    \bigskip
	\begin{footnotesize}
	    This figure depicts the difference between central (red) and peripheral (brown) nodes in a graph.
    \end{footnotesize}
    \label{fig1}
\end{figure}

Another important centrality measure is the \quotes{closeness}, which measures how long it takes for information to spread from one node to the next. Specifically, closeness is defined as the reciprocal of the \quotes{farness}, \textit{i.e.} the sum of distances of one node with respect to all other nodes. The \quotes{betweenness centrality} of a node measures how often a node stands in the shortest path between a pair of other nodes (see, \textit{e.g.}, Bavelas; 1948, Saxena and Iyengar; 2020, and Freeman; 1978)\citealp{Bavelas1948,SaxenaIyengar2020W,Freeman1978}).

Another strand of research focuses on the top-K shortest path identification in a complex network, a topic less tackled by the literature than centrality. To rank nodes, one must compute the centrality of all nodes and compare them to extract the rank, which is not always feasible due to the size of the network. To overcome this problem Saxena and Iyengar (2017) attempt to estimate the global centrality of a node without analyzing the whole network\cite{SaxenaIyengar2017W}. Similarly, Bavelas (1948) develops a structural centrality measure in the context of social graphs\cite{Bavelas1948}. Other centrality concepts include the eigenvector, Katz, or PageRank centralities\cite{Bonacich1972,PageBrinMotwaniWinograd1999,Katz1953}. Finally, Freeman (1978) creates a formal mathematical framework for centrality, which includes degree, closeness, and betweenness and advocates for the combination of different kinds of centrality measures\cite{Freeman1978}.

\subsection{Page Rank}
Page, Brin, Motwan, and Winograd (1999) develop the PageRank algorithm\cite{PageBrinMotwaniWinograd1999}. Its primary goal is to rank web pages objectively, a challenge with the fast-growing web. PageRank assigns a score to each web page based on its relations with other web pages in the graph. Other fields have benefited PageRank providing modifications and improvements. Xing and Ghorbani (2014) extend the algorithm and propose the weighted PageRank (WPR)\cite{XingGhorbani2004}. This algorithm assigns larger rank values to more important pages, instead of dividing the rank evenly among its outlink pages.\footnote{Given a web page W, an inlink of W is a link of another web page that includes a link pointing to W. An outlink of W is a link appearing in W, which points to another web page.} Each outlink page gets a value proportional to its popularity, taking into account the links weights. On caveat of PageRank and its variants is that they do not consider n-partite structures, yet, web pages can all be linked to one another. Bi-partite networks address this issue and capture this complexity, among other structures. 

\subsection{Bi-partite networks}

\begin{figure}[h!]
    \centering
    \includegraphics[scale=1.6]{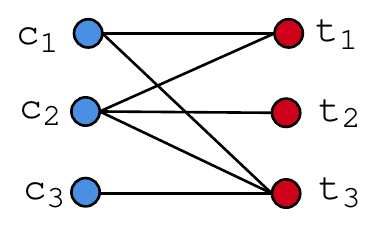}
    \includegraphics[scale=0.6]{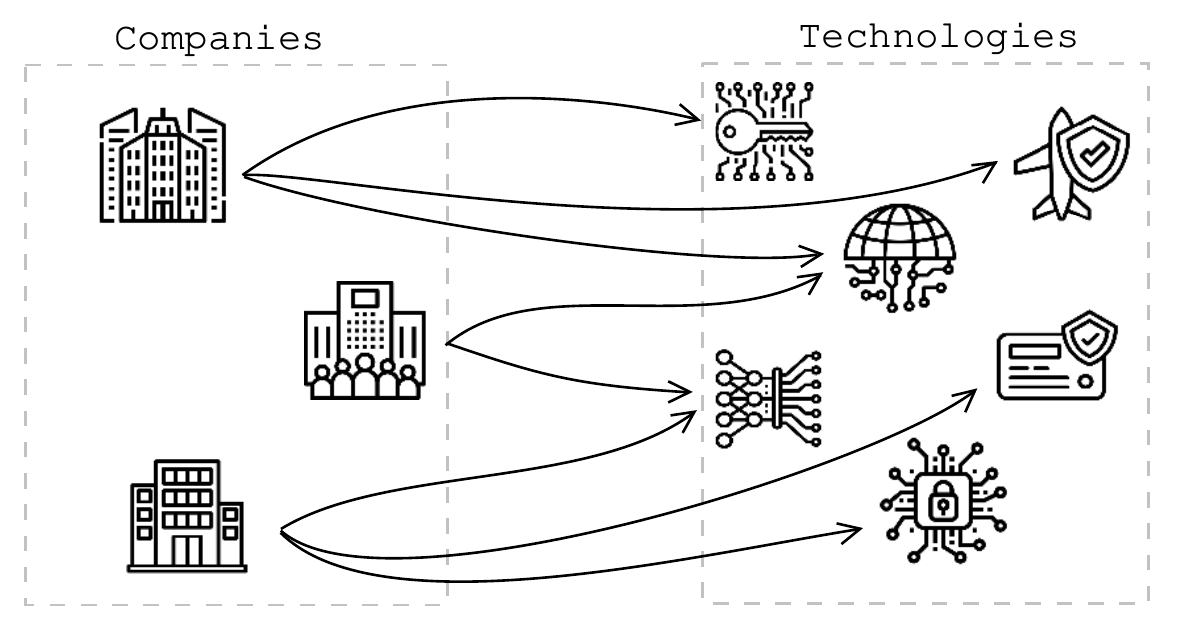}
    \caption{Bi-partite structure of companies and technologies}
    \bigskip
	\begin{footnotesize}
	    The left panel depicts a typical bi-partite structure. The right panel provides an illustration of this structure with companies (layer 1) and technologies (layer 2).\bigskip
    \end{footnotesize}
    \label{fig2}
\end{figure}

Networks are a fundamental tool to capture the relations between entities. Graphs ($G$) are composed by vertices ($V$) and edges ($E$), and we denote $G=(V,E)$. We build links and mathematically analyze many properties of the whole system and of singular entities. To graphically represent a real system, we synthesize its information into a simple graph framework. This simplification generates an information loss in the modelling process. Simple network structures might discard important information about the structure and function of the original system \cite{KurantThiran2006}. As a consequence, the failure of a very small fraction of nodes in one network may lead to the complete fragmentation of a system\cite{BuldyrevParshaniPaulStanleyShlomo2010}. To solve the problem, extensions to the simple structure $G=(V,E)$ are added and yield graphs with more powerful features. For instance, in case vertices are connected by relationships of different kinds, Battiston, Nicosia, and Latora (2014) advocate to work with multiplex networks, \textit{i.e.} networks where each node appears in a set of different layers, and each layer describes all the edges of a given type \cite{BattistonNicosiaLatora2014}. When it is possible to distinguish the nature of the edges, multiplex networks are an effective approach, which starts from embedding the edges in different layers according to their type. However, even if we have two kinds of nodes, the nature of the edges is unique. Therefore, a more suitable approach is bi-partite networks. Bi-partite networks are for instance, a good way to describe the technological and business landscape. In Figure \ref{fig2}, we depict two sets of nodes, companies and technologies, which are interconnected but do not present edges within the same set.

There are multiple adaptations of the PageRank algorithm to bi-partite structures\cite{BenziEstradaKlymko2013, DonatoLauraLeonardiMillozzi2004, TuJiangSongZhang2018, KleinMaillartChuang2015}. In particular, Benzi, Estrada, and Klymko (2015), Donato, Laura, Leonardi, and Millozzi (2004), and Tu, Jiang, Song, and Zhang (2018) extend the PageRank algorithm to multiplex networks. They assume that only some clusters of the graph are multiplex networks and extend the PageRank algorithm only to analyze the sub-graph centrality. Bi-partite networks are used to transform directed into undirected networks with twice the number of vertices.

Klein, Maillart, and Chuang (2015) extend PageRank in the Wikipedia editors and articles context \cite{KleinMaillartChuang2015}. The application of this algorithm to the case of interactions between companies and technologies is straightforward. A major benefit of this approach is that it starts from an unweighted graph, linking authors and articles. They develop a recursive algorithm in which the two entities contribute to the quality (for articles) or the expertise (for authors) of each other. They develop a bi-partied random walker by building the adjacency matrix $M_{e,a}$ that takes value 1 if editor $e$ has edited article $a$ and 0 otherwise, which tracks all the editors' contributions. They obtain $M_{e,a} \in \mathbb{R}^{n_e, n_a}$, where $n_e$ and $n_a$ are the number of editors and articles. They sort editors by the number of articles' contributions and assign a contribution (quality) value to each editor (article) based on their degree. The expertise $w_e^{0}$ (quality $w_a^{0}$) is given by the number of articles (editors) they have worked on (have received modifications).

The second part of the algorithm follows a Markov process in its iterations. The step $w^{n}$ ($w^{n} = w^{n}(\alpha, \beta)$) only depends on information available at $w^{n-1}$. At each step, the algorithm incorporates information about the expertise of editors and the quality of articles, within the bi-partite network structure. The process is a random walker with jumps, whose transition probability is zero in the case $M_{e,a}=0$. Next, the authors define two variables for the transition probability, $G_{e,a}(\beta)$ and $G_{e,a}(\alpha))$. $G_{e,a}(\beta)$ represents the probability of jumping from article $a$ to editor $e$ and $G_{a,e}(\alpha)$ represents the probability to jump from an editor to an article. Both parameters depend on initial conditions and the selection of optimal parameters is done through a grid search that maximizes the Spearman rank-correlation between the rank given by the model and a ground-truth metrics obtained independently. Finally Klein et al. (2015) observe a \quotes{less-is-more} situation since the presence of too many editors working on an article is detrimental to its quality. Studying different categories of Wikipedia articles they find $\alpha$ to remain constant, while $\beta$ varies significantly across categories.

Estimating the global rank of a node starting from local information and centrality measures is still an open research question in many sectors \cite{SaxenaIyengar2020W}. In particular, no research to our knowledge use this approach for investment's decision and portfolio optimization. Yet, this approach could help overcome the limitations of standards financial models in private equity, in which the network structure is easily obtainable, whereas the cash flow process is not.

\subsection{Private equity valuation}
Private firms are not required to publicly disclose their financial statements, which makes it difficult to measure their past performance and estimate their expected returns, without insider information. Moreover, since they are not listed on exchanges we do not observe expectations of market participants. Thus, standard asset pricing methods fall short in this context. Similarly, private equity analysts must rely on insider or private information to value private firms. These valuations generally occur around a financing round and impose to take into account the capital dilution to compute realized returns on the firm\cite{GornallStrebulaev2020}.

One approach that attempts to overcome these limitations and estimate the expected returns and risk in venture capital is Cochrane (2005)\cite{Cochrane2005}. He uses a maximum likelihood estimation method to obtain these values at the market and sector levels, such as healthcare, biotechnology, technology companies, and retail services. He finds a mean arithmetic return of 59\%, an alpha of 32\%, a beta of 1.9 and a volatility of 86\% (equivalent to a 4.7\% daily volatility). Given that the distribution of returns is heavily positively skewed in venture capital, he adopts a logarithmic model that also accounts for the inherent selection bias of this asset class.\footnote{Most of venture capital data is private and available data are more often related to successful firms than under-performing ones.} Ewens (2009) updates this method on returns computed from financing round to the next\cite{Ewens2009}. He adopts a three-regime mixture model (failure, medium returns, and \quotes{home-runs}). He also corrects for the selection bias and obtains an alpha of 27\% and a beta of 2.4. He finds that 60\% of all venture capital investments have a negative log return. Altogether the results are similar, venture capital investments exhibit positive alpha, large beta, and a high volatility. Other attempts to evaluate the market parameters of the venture capital asset class includes Korteweg and Nagel (2016) and Moskowitz and Vissing-Jorgensen (2010) with results in line with the aforementioned studies\cite{KortewegNagel2016,MoskowitzVissing-Jorgensen2002}.

Another strand of research attempts to index and benchmark the private equity market. Peng (2001)  builds a venture capital index from 1987 to 1999 from about 13,000 financing rounds targeting over 5,600 firms\cite{Peng2001W}. He addresses the problems of missing data, censored data, and sample selection using a re-weighting procedure and method of moment regressions. From the index perspective, the results are qualitatively the same, that is, high and volatile returns to venture capital (average return of 55.18\% per year). He finds his index to display a much higher volatility than the S\&P 500 and NASDAQ indices and high exposure to these indices (betas of 2.4 and 4.7, respectively). Other venture capital indices construction includes Hwang, Quigley, and Woodward (2005), Schmidt (2006), and Cuming, Has and Schweizer (2013), all obtaining results on par with the aforementioned studies.

One limitation of the above studies is that they only estimate these parameters at an aggregate level. An investor could form her investment's decisions and portfolio choices segregating among sectors, but not obtain the actual firms' parameters. One exception is Schwartz and Moon (2000), who provide an approach based upon real-options theory to price individual firms\cite{SchwartzMoon2000}. However, this method requires the observations of cash flows and they only provide one calibration example with Amazon. Thus, there remains a caveat in methodology to help investors' forming optimal decisions using all the information available. Given the recent venture capital boom, Zhong, Chuanren, Zhong, and Xiong (2018) advocates for the use of quantitative methodologies of screening and evaluation\cite{ZhongChuanrenZhongXiong2018}. However, there is a clear research gap on methodologies enabling to value early stage companies and form optimal portfolios. These methodologies either only enable to value a sector, instead of a specific company, or require the use of cash flows that are unobservable. Non-financial features and relations between companies and technologies are instead numerous and easily observable\cite{DalleBestenMenon2017W}. We thus formulate our hypotheses as follows,

$H_1$ Using a bi-partite network structure allows to create an algorithm that rank companies based on their links with technologies.

$H_2$ This algorithm and its ranking may be improved and tilted towards investors' preferences.

$H_3$ The performance of the ranking is independent from the sector considered.

\section{Data and methods}\label{sec3}

\subsection{Data}
We use Crunchbase data.\footnote{Crunchbase website: \url{https://www.crunchbase.com/}} Crunchbase is a commercial database that provides access to financial and managerial data on private and public companies globally. It was created in 2007 by TechCrunch, which is a source of information about start-up activities and their financing within and across countries. This database has been largely adopted by both academics and industry practitioners\cite{Besten2021}. It is also used by international organizations such as the OECD\cite{DalleBestenMenon2017}. 

Crunchbase is made of data collected with a multifaceted approach that combines crowd-sourcing (through venture programs or direct contributions), machine learning, in-house processing, and aggregation of third-party providers data. Crunchbase updates and revises data on a daily basis, which is organized into several entities such as organizations, people, events, acquisitions, or IPOs. The primary focus of Crunchbase is the technology industry, although it also includes data on other sectors.

Data can be accessed in two ways: using an API or downloading a *.csv file directly from the Crunchbase website. Data is split in several databases depending on their type. We provide a non-exhaustive list in Table \ref{tab1}.\footnote{Crunchbase daily CSV export from \url{https://data.crunchbase.com/docs/daily-csv-export}. Data downloaded on April 28, 2021.}

\begin{table}[h!]
    \centering 
    \begin{tabular}{ll}
        Field name                  & Description\\ \cmidrule(lr){1-1} \cmidrule(lr){2-2}
        
        organizations               & Organisation profiles\\ 
        organization\_desc          & Long descriptions of organisation profiles\\
        acquisitions                & List of all acquisitions\\
        org\_parents                & Map between parent organisations and subsidiaries\\
        ipos                        & Detail for each IPO\\
        people                      & People profiles\\
        people\_desc                & Long descriptions for people profiles\\
        degrees                     & Detail of people's education background\\
        jobs                        & List of all jobs and advisory roles\\
        investors                   & Active investors (organisations and people)\\
        investments                 & All investments\\
        investment\_partners        & Partners responsible for their firm's investments\\
        funds                       & Details of investments funds\\
        funding\_rounds             & Details for each funding round in the dataset\\
        events                      & Event details\\
        event\_appearances          & Event participation details\\
    \end{tabular}
    \caption{Crunchbase files description}
    \bigskip
    \begin{footnotesize}
	    This table reports the main fields available from Crunchbase.\bigskip
    \end{footnotesize}
    \label{tab1}
\end{table}

\begin{figure}[h!]
    \centering
	\includegraphics[scale=0.5]{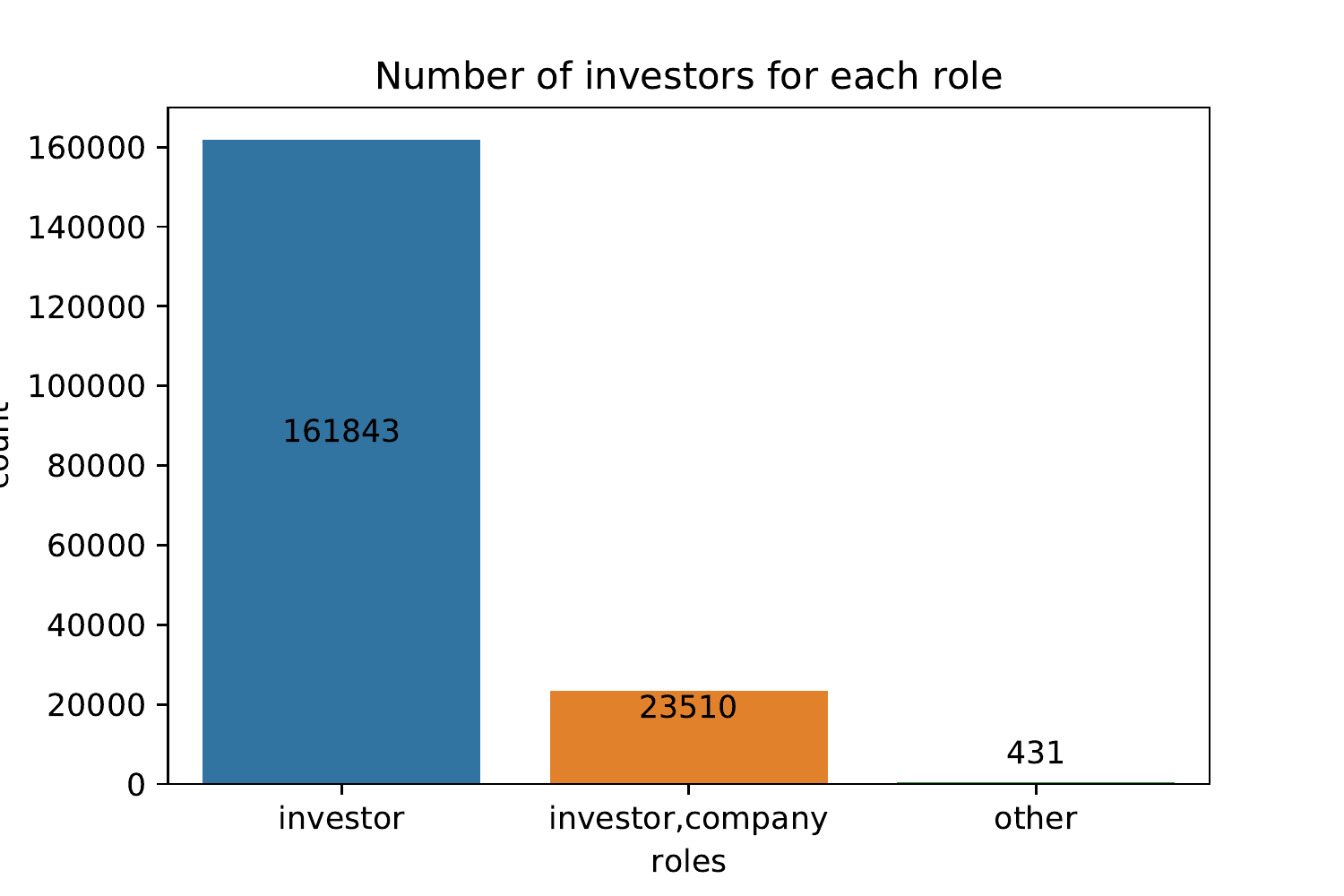}
	\includegraphics[scale=0.5]{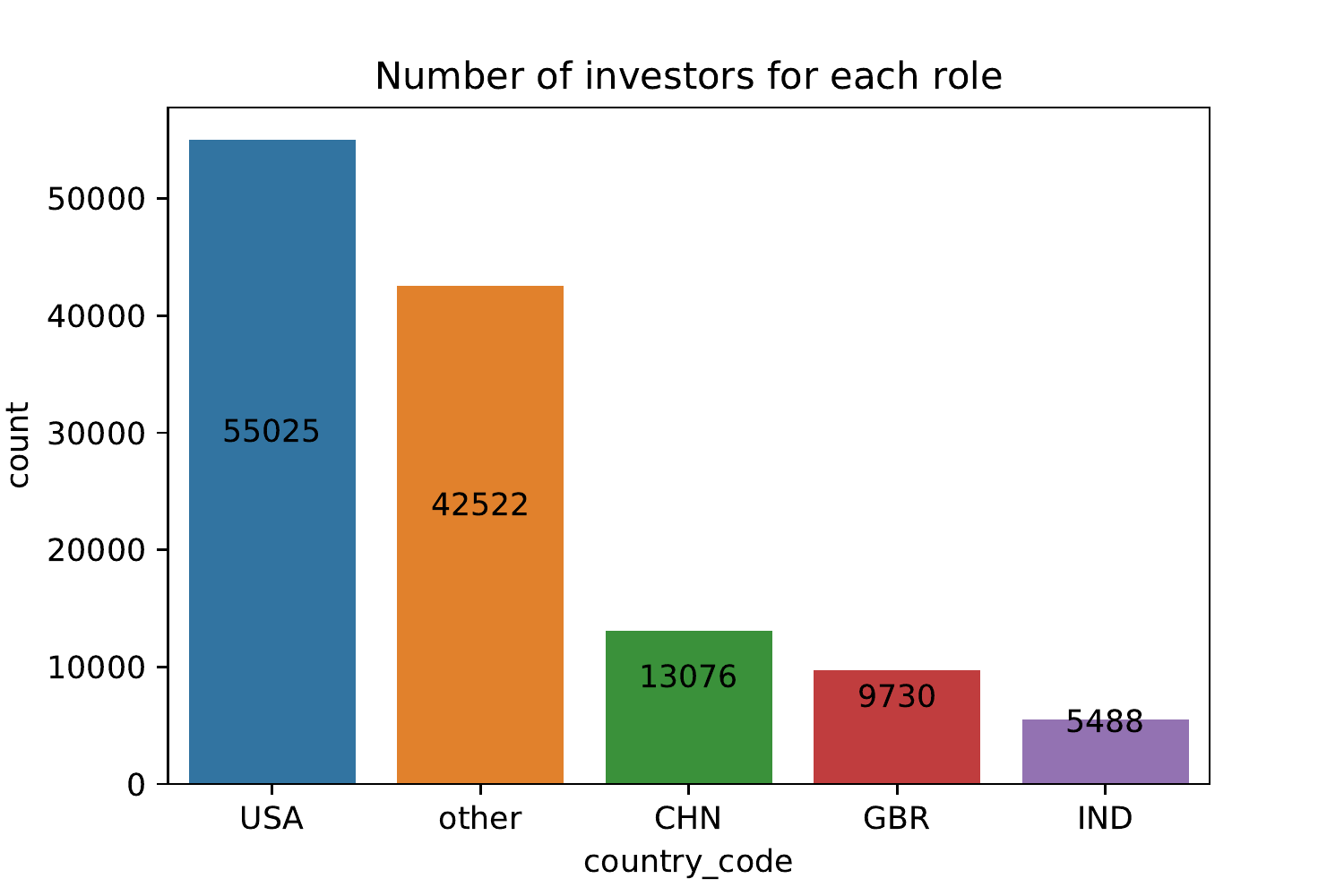}
	\caption{\textbf{Summary statistics of investors in Crunchbase data}}
	\bigskip
	\label{fig3}
	\begin{footnotesize}
		The left panel depicts the distribution of investors according to their types. The right panel depicts the number of investors per country.\bigskip
	\end{footnotesize}
\end{figure}

We first analyze the Crunchbase dataset dedicated to investors. With a total of 185,784 investors divided into 78,001 (41.98\%) organisations and 107,783 (58.02\%) persons, there are more investors than target companies. Figure \ref{fig3} shows that the majority of investors are pure investors (87.11\%). Some organisations are both investee and investor (12.65\%). The remainder of the sample  are typically universities. Crunchbase ranks the top 1,000 investors through its proprietary algorithm. Figure \ref{fig3} indicates that the majority of investors are located in the USA (29.62\%). In particular, there is a wide gap between the first and second country, China, where 7.04\% of investors are located.\newpage

%

\subsection{Methodology}

\subsubsection{Adaptation of the work by Klein et al. (2015).}

In this research we use a bi-partite network that describes the relations among companies and the technologies they are involved in. Figure \ref{fig2} describes the typical bi-partite network structure. This structure benefits from advances in fields such as network theory, Markov chains, and machine learning. We adapt the recursive algorithm with the method of Hidalgo, Hausmann, and Dasgupta (2009)\cite{HidalgoHausmannDasgupta2009}. We expect the resulting rank to incorporate the positive influence of well-established companies on technologies and, at the same time, the positive impact of new companies that explore new fields. We build the adjacency matrix $M^{CT}_{c,t}\in \mathbb{R}^{n^c, n^t}$, which takes value 1 if a company $c$ works on a technology $t$ and 0 otherwise. $n^c$ and $n^t$ represent the number of companies and technologies. We assume that well-established companies have more means to diversify their expertise and therefore, that an entity has a relatively high number of neighbours\cite{CanitoRamosMoroRita2018, GoldMalhotraSegars2001}. Thus, we initialize the algorithm with the degree, \textit{i.e.} counting the neighbours, of each entity,
\begin{equation}\label{eq1}
    \begin{cases}
    w_c^{0} &= \sum_{t=1}^{n^t} M^{CT}_{c,t} = k_c \\
    w_t^{0} &= \sum_{c=1}^{n^c} M^{CT}_{c,t} = k_t
    \end{cases}
\end{equation}
The algorithm is a \quotes{random walker} that incorporates information about company expertise and technology relevance at each step. The transition probabilities, $G_{c,t}$ and $G_{t,c}$, describe the extent to which the entities weights change over the iterations. If the relation between $c$ and $t$ increases (decreases) the value, the entity weight increases (decreases) in proportion with the transition probabilities. We define $G_{c,t}$ and $G_{t,c}$,
\begin{equation}\label{eq2}
    \begin{cases}
    G_{c,t}(\beta) &= \frac{M^{CT}_{c,t}k_c^{-\beta}}{\sum_{c'=1}^{n^c} M^{CT}_{c',t}k_c'^{-\beta}}  \\
    G_{t,c}(\alpha) &= \frac{M^{CT}_{c,t}k_t^{-\alpha}}{\sum_{t'=1}^{n^t} M^{CT}_{c,t'}k_t'^{-\alpha}},
    \end{cases}
\end{equation}
where $\alpha$ and $\beta$ inform how coordination generates value. Next, we define the recursive step,
\begin{equation}\label{eq3}
    \begin{cases}
    w_c^{n+1} &= \sum_{t=1}^{n^t} G_{c,t}(\beta) w_t^{n}  \\
    w_t^{n+1} &= \sum_{c=1}^{n^c} G_{t,c}(\alpha) w_c^{n}
    \end{cases}
\end{equation}
As in PageRank, the recursion ends when the algorithm converges. Our algorithm allows to consider the market complexity and feedback loops (investments' impact on companies and on technologies). We discuss this feature and the optimization of $\alpha$ and $\beta$ after the addition of exogenous factors.

\subsubsection{Inclusion of exogenous factors}

We include exogenous factors as ground truth in the parameters' calibration step. This allows to keep the algorithm tractable, while letting it capture the technological structure. We use this ground truth to compute the Spearman correlation, $\rho_c$ for companies and $\rho_t$ for technologies. Because $\rho_c$ and $\rho_t$ depend on $\alpha$ and $\beta$ [see Eq. (\ref{eq2})], we find the parameters that maximize these correlations,
\begin{equation}\label{eq4}
    \begin{cases}
    (\alpha^*, \beta^*) = \argmax_{\alpha, \beta} \rho_c(\alpha, \beta) \\
    (\alpha^*, \beta^*) = \argmax_{\alpha, \beta} \rho_t(\alpha, \beta),
    \end{cases}
\end{equation}
and we solve this optimization problem with a grid search. Eq. (\ref{eq4}) shows that parameters depend on both companies and technologies. This dependence enables to create the structure of the bi-partite graph. To obtain the correlation between the TechRank score, which assigns a weight $w_c$ ($w_t$) to each company (technology) and the ground truth evaluation, which assigns $\hat{w_c}$ ($\hat{w_t}$) to each company (technology), we normalize both TechRank results and the exogenous measure in the same range $[0,1]$.

Investors use the entities' features to select companies and the investment amount they want to allocate. We suppose that an investor has $n^{(C)}$ features to pick from, denoted as $f^{(C)}_1,\ldots,f^{(C)}_{n^{(C)}}$, where $C$ ($T$) represents the association with the companies (technologies). Each feature $f^{(C)}_i$ is associated with a percentage of interest $p^{(C)}_i$ and $\sum_{i=0}^{n^{(C)}} p^{(C)}_i = 1$. For instance, if a company's features are the amount of previous investments and its geographical proximity to the investor, $n^{(C)}=2$. An investor may then decide to be interested at 80\% in the first feature and at 20\% in the second, by selecting $p^{(C)}_1 = 0.8$ and $p^{(C)}_2 = 0.2$. Investors may also be pushed back by a feature, in which case we multiply it by -1. We define all notations in Table \ref{tab2}.\newpage
\begin{table}[h!]
\centering
\resizebox{0.8\linewidth}{!}{\begin{tabular}{ccl}
Variable & $\in$ & Description     \\ \cmidrule(lr){1-1} \cmidrule(lr){2-2} \cmidrule(lr){3-3} 
$n^{(C)}$        & $\mathbb{N}$ & Number of external features available for companies.        \\ 
$n^{c}$          & $\mathbb{N}$ & Number of companies.                                        \\
$n^{(T)}$        & $\mathbb{N}$ & Number of external features available for technologies.     \\ 
$n^{t}$          & $\mathbb{N}$ & Number of technologies.                                     \\ 
$p^{(C)}_i$      & $[0,1]$      & Percentage of interest in the company preference number $i$.\\ 
$p^{(T)}_j$      & $[0,1]$      & Percentage of interest in the technology preference number $j$.\\ 
$f^{(C)}_i$      & $\mathbb{R}^{n^c}$ & Vector of factors associated to the company preference number $i$.\\ 
$f^{(T)}_j$      & $\mathbb{R}^{n^t}$ & Vector of factors associated to the technology preference number $j$.\\

$n^{i}$              & $\mathbb{N}$        & Number of investors.\\ 
$M^{CT}$             & $\mathbb{R}^{n^c \cdot n^t}$ & Adjacency matrix of the C-T bipartite network\\
$M^{IC}$             & $\mathbb{R}^{n^i \cdot n^c}$ & Adjacency matrix of the I-C bipartite network\\
$\gamma^{i,c}_t$     & $\mathbb{R}$ & Amount in funding round between $c$ and $i$ at time $t$\\ 
$e^{IC}$             & $\mathbb{R}^{n^i \cdot n^c}$ & Total amount of investment between each investor to each company\\ 
$e^{C}$              & $\mathbb{R}^{n^c}$ & Total amount of investments toward each company\\ 
$e^{T}$              & $\mathbb{R}^{n^t}$ & Total amount of investments toward each technology\\ 
$e^{C}_{\text{max}}$ & $\mathbb{R}$  & Maximum amount of total investments among all the companies\\ 
$e^{T}_{\text{max}}$ & $\mathbb{R}$ & Maximum amount of total investments among all the technologies\\ 
$f^{C}_c$            & $[0,1]$ & Factor related to previous investments into the company number $c$\\
$f^{T}_t$            & $[0,1]$& Factor related to previous investments into the technology number $t$ \\ 
\end{tabular}}
\caption{Variable definitions}
\bigskip
\label{tab2}
\begin{footnotesize}
		This table presents the variable definitions used throughout the article.\bigskip
	\end{footnotesize}
\end{table}

We convert quantitative and qualitative properties into a number $f^{(C)}_i \in [0,1]$. Once we have created all the factors $f^{(C)} = f^{(C)}_1,\ldots,f^{(C)}_{n^{(C)}}$, the exogenous evaluation $\hat{w_c}$ is given by,
\begin{equation}\label{eq5}
    \hat{w_c} = \sum_{i=1}^{n^{(C)}} p^{(C)}_i f^{(C)}_i = p^{(C)} \cdot f^{(C)}.
\end{equation}
Considering $\sum_{i=0}^{n^{(C)}} p^{(C)}_i = 1$ and that $f^{(C)}_i \in [0,1]$ for each company $i$, we have $\hat{w_c} \in [0,1]$. The same holds for $\hat{w_t}$. Finally, we have
\begin{equation}\label{eq6}
    \begin{cases}
    \hat{w_c} =  p^{(C)} \cdot f^{(C)}\\
    \hat{w_t} =  p^{(T)} \cdot f^{(T)}\\
    \sum_{i=0}^{n^{(C)}} p^{(C)}_i = 1 \\
    \sum_{i=0}^{n^{(T)}} p^{(T)}_i = 1,
    \end{cases}
\end{equation}
where $n^{(C)}$ ($n^{(T)}$) is the number of the company- (technology-)related features and $f^{(C)} = (f^{(C)}_1,\ldots,f^{(C)}_{n^{(C)}})$. To select the features, we use Crunchbase data about companies and investors (see in Table \ref{tab1}).

\subsection{Previous Investments}

We assume that previous investments is an essential factor to evaluate companies. Investors may be willing to invest in companies which have already received capital or look for higher returns, targeting newer firms. 

To compute this factor, we use the Crunchbase field \quotes{funding\_rounds}, which reports the amount of all funding rounds from an investor $i$ to a company $c$. We capture this structure with another bi-partite network that describes the links among investors (I) and companies (C). In this case, we weight the edges by the sum of all previous investments from investor $i$ to company $c$, until the current period ($\mathcal{T}$), and compute the adjacency matrix $M^{IC}$. We define the amount of a single investment from $i$ to $c$ at time $t$ by $\gamma^{i,c}_{t}$. The weight of the edge $i-c$ is given by $e_{i,c} = \sum_{t=0}^{\mathcal{T}} \gamma^{i,c}_{t}$ (see Table \ref{tab2}). We then sum the contribution of all investors to find the attribute $f^{C}_c \in [0,1]$ for a company $c$.\footnote{Note that here, $f_c^{C}$ represents the factor related to a company.} Next, we normalize and divide all investments by the maximum investment. 

\begin{figure}[h!]
    \centering
    \includegraphics[scale=1.5]{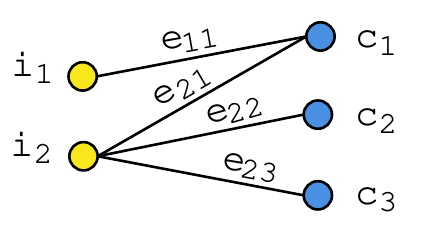}
    \caption{Investors-companies bi-partite network}
    \bigskip
    \label{fig4}
    \begin{footnotesize}
		This figure depicts a stylized bi-partite network between investors and companies.\bigskip
	\end{footnotesize}
\end{figure}

Figure \ref{fig4} depicts the investment structure as an example. We consider two investors $i_1$ and $i_2$ and three companies $c_1$, $c_2$, and $c_3$. We compute the maximum $e_{\text{max}}$ as $\max\{e_{11}+e_{21}, e_{22}, e_{23}\}$ and the features related to the investments for each company as,
\begin{equation}\label{eq7}
    \begin{cases}
    f^{C}_1 = \frac{e_{11}+e_{21}}{e_{\text{max}}}\\
    f^{C}_2 = \frac{e_{22}}{e_{\text{max}}}\\
    f^{C}_3 = \frac{e_{23}}{e_{\text{max}}},
    \end{cases}
\end{equation}
where $n^{i}$ ($n^{c}$) is the total number of investors (companies). Generalizing, we get,
\begin{equation}\label{eq8}
    \begin{cases}
    e^{IC}_{i,c} = \sum_{t=0}^{\mathcal{T}} \gamma^{i,c}_{t}  \hspace{1.4cm} \forall i, c  \\
    e^C_{c} = \sum_{i=1}^{n^{i}} e_{i,c} M^{IC}_{i,c}  \hspace{0.9cm} \forall c  \\
    e_{\text{max}} = \underset{c}{\max{}} e^C_c \\
    f^{(C)}_c =  e^C_{c} / e_{\text{max}},
    \end{cases}
\end{equation}
for each $c \in {1,\ldots, n^{c}}$. We present the corresponding algorithm in the Appendix \ref{algo1}. With Eq. (\ref{eq8}), for each company we have a factor between 0 and 1 that summarizes the amount of previous investments.

\begin{figure}[h!]
    \centering
    \includegraphics[scale=1.2]{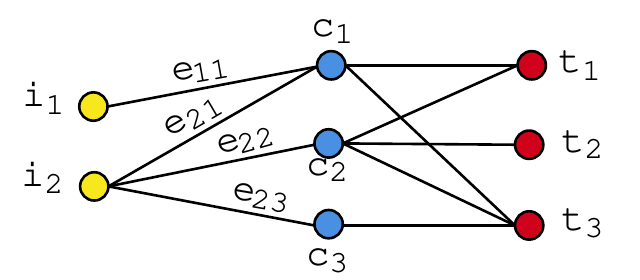}
    \includegraphics[scale=0.6]{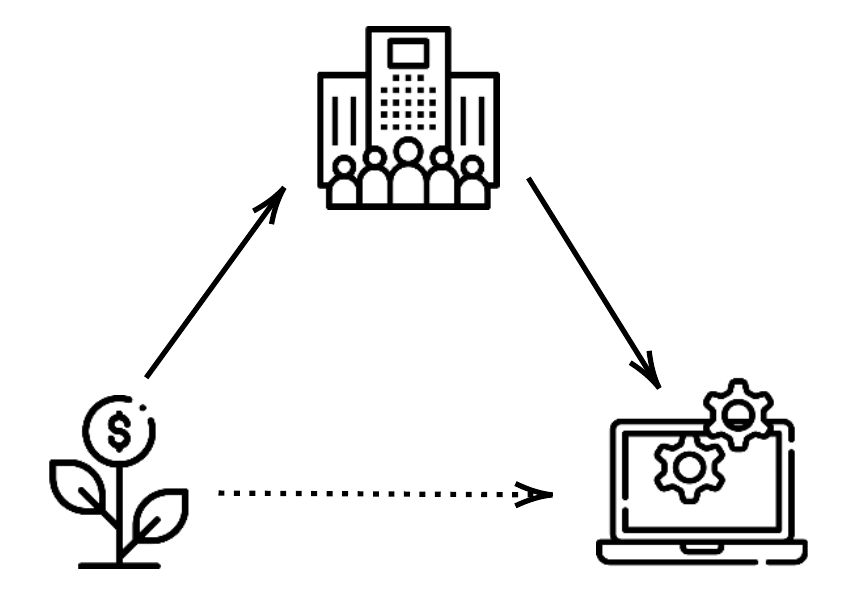}
    \caption{Tripartite structure of investors, companies, and technologies}
    \bigskip
    \label{fig5}
    \begin{footnotesize}
        The left hand side figure depicts a typical tri-partite structure. The right-hand side provides an illustration of this structure with investments (layer 1), companies (layer 2), and technologies (layer 3).\bigskip
    \end{footnotesize}
\end{figure}
We link the two bi-partite structures investment-companies and companies-technoéogies to obtain an I-C-T tri-partite structure depicted in Figure \ref{fig5}. This structure allows to assign some features to technologies from companies (direct link) or investors (indirect link). Thus, we can find the amount of previous investments on a technology through companies' funding rounds. The previous investments' factor for technology is given by,
\begin{equation}\label{eq9}
    \begin{cases}
    e^{(I,C)}_{i,c} = \sum_{t=0}^{(T)} \gamma^{t}_{i,c}  \hspace{0.6cm}\forall i, c \\
    e^C_{c} = \sum_{i=1}^{n^{i}} e_{i,c}  \hspace{1.1cm}\forall c  \\
    e^T_{t} = \sum_{c=1}^{n^{c}} e_{c}  M^{CT}_{c,t}  \\
    e_{\text{max}} = \underset{t}{\max{}} e^T_t\\ 
    f^{(T)}_t =  e^T_{t} / e_{\text{max}}
    \end{cases}.
\end{equation}

We provide the algorithm of this methodology in the Appendix \ref{algo2}.

\subsection{Location}
The second feature we consider is the distance between investors and companies' locations. We retrieve the addresses of firms and investors from Crunchbase ($c\_address$) and map them to geographic coordinates. We compute the Haversine approximation to measure the distance. We detail the Haversine approximation in the Appendix \ref{app4}. Investors may prefer short-distance investments or places with high potential. If they face some investment's restrictions, we filter the companies based on the criteria before applying the algorithm. Otherwise, we add a distance factor to the algorithm.

We use the Haversine distance $h$ to obtain a factor $f_c^{(C)} \in [0,1]$ for each company. We consider the distance $h_{i,c}$ between the company $c$ and the investor $i$. We assume that the factor is the proximity so that $f_c^{(C)}$ tends to one as the distance decreases,
\begin{equation}\label{eq10}
    f_c^{(C)} \rightarrow 1  \hspace{0.5cm} \text{when} \hspace{0.5cm} h_{i,c} \rightarrow 0.
\end{equation}
To compute $f_c^{(C)}$, we first find $h_{i,c}$ for each company and identify the maximum distance $h_{\text{max}}$ among all companies. We normalize by the maximum to obtain a distance that lies in the $[0,1]$ range with, $f_c^{(C)} = 1 - h_{i,c}/h_{\text{max}}$, so that a distance of zero corresponds to a value of $f_c^{(C)} = 1$. We report the algorithm in the Appendix \ref{algo3}.
We implement the algorithm and run the experiments using Python and the NumPy, Pandas, NetworkX, Matplotlib, and Seaborn libraries.
\section{Results}\label{sec4}
\subsection{Cybersecurity field}

We select all the companies whose description contains at least two cybersecurity-related terms and obtain 2,429 companies and 477 technologies.\footnote{The word list is in the Appendix \ref{app2}} Figure \ref{fig6} display the structure of the bi-partite network between technologies and companies.\newpage

\begin{figure}[h!]
    \centering
    \includegraphics[scale=0.27]{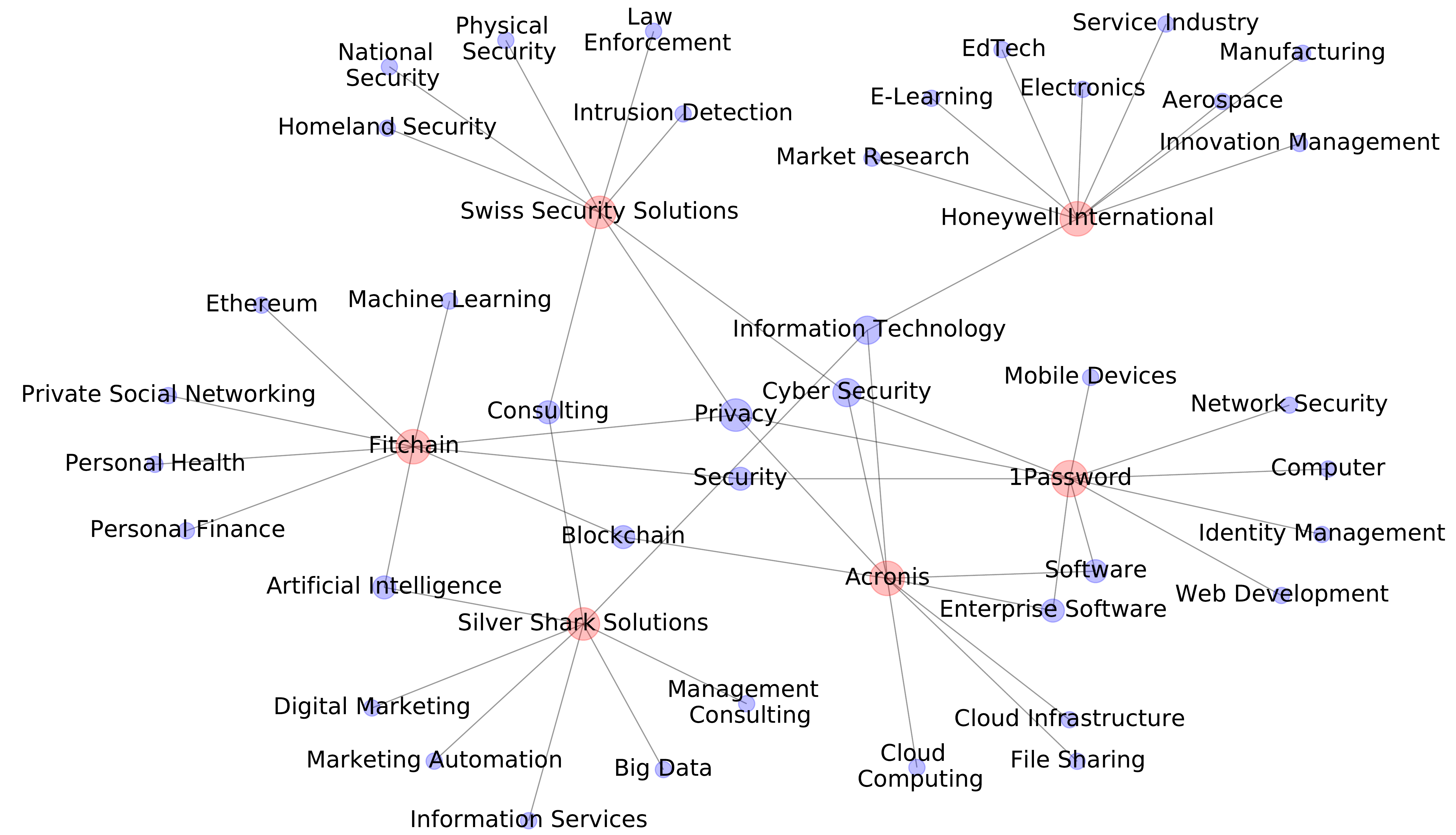}
    \caption{Bi-partite network of cybersecurity companies}
    \bigskip
    \label{fig6}
    \begin{footnotesize}
        This figure describes the bi-partite network of a subset of cybersecurity companies (red nodes) and the technologies they are involved in (blue nodes). The nodes size represents the number of neighbours.\bigskip
    \end{footnotesize}
\end{figure}

\begin{figure}[h!]
    \centering
    \includegraphics[scale=0.34]{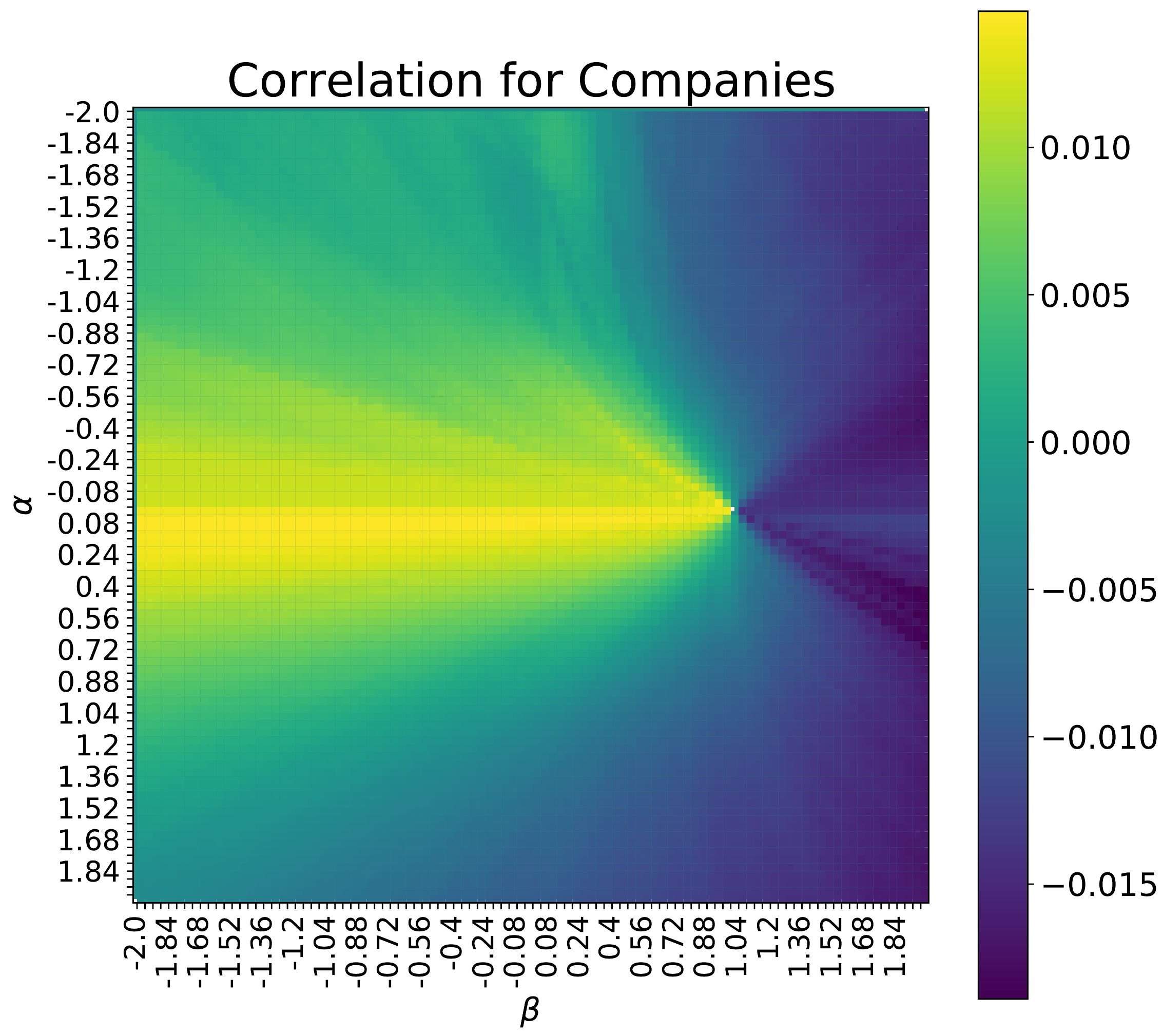}
    \includegraphics[scale=0.33]{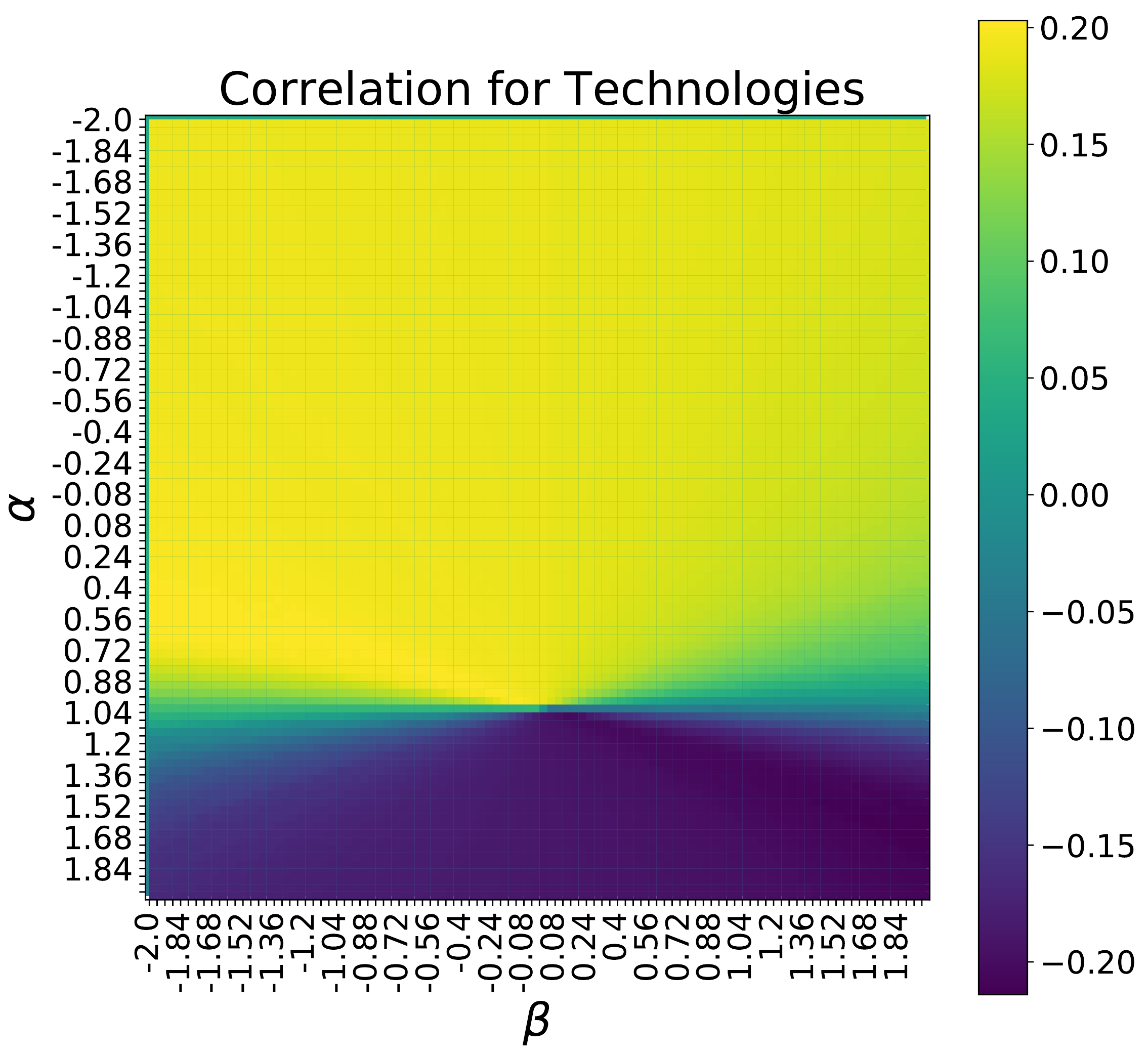}
    \caption{Grid search of parameters $\alpha$ and $\beta$}
    \bigskip
    \label{fig7}
	\begin{footnotesize}
		This figure displays the results of the grid search for parameters $\alpha$ and $\beta$ for 2,429 companies and 477 technologies in cybersecurity when investors preferences are fully set in previous investments.\bigskip
	\end{footnotesize}
\end{figure}

We assume that investors are only interested in previous investments, both for technologies and companies. We examine how the parameters' calibration step changes when we change the investors' preferences using a smaller sample of companies. Figure \ref{fig7} shows the optimization in which the correlations $\rho_c$ and $\rho_t$ change according to $\alpha$ and $\beta$. In Table \ref{tab3} We identify the optimal $\alpha^*$ and $\beta^*$ to be of 0.04 and -1.88 for companies and 0.48 and -2.00 for technologies, respectively. Next, we plug these values in the recursive algorithm.

\begin{table}[h!]
    \centering
    \begin{tabular}{cccccc}
       \multicolumn{3}{c}{Companies} & \multicolumn{3}{c}{Technologies}\\\cmidrule(lr){1-3} \cmidrule(lr){4-6} 
       Number & $\alpha^*$ & $\beta^*$ & Number & $\alpha^*$ & $\beta^*$\\\cmidrule(lr){1-1} \cmidrule(lr){2-2} \cmidrule(lr){3-3} \cmidrule(lr){4-4} \cmidrule(lr){5-5} \cmidrule(lr){6-6}
        10          & -0.36        & 1.92        & 26          & -2.00        & 0.00        \\
        100         & -0.04        & 0.92        & 134         & 0.52         & -1.04       \\
        499         & -0.08        & 0.88        & 306         & 0.68         & -1.36       \\
        997         & -0.12        & 0.80        & 371         & -2.00        & 0.00        \\
        1,494       & -0.12        & 0.80        & 416         & 0.92         & -0.12       \\
        1,990       & -0.04        & 0.92        & 449         & 0.56         & -2.00       \\
        2,429       & 0.04         & -1.88       & 477         & 0.48         & -2.00      
    \end{tabular}
    \caption{Optimal parameters in cybersecurity}
    \bigskip
    \label{tab3}
    \begin{footnotesize}
        This table reports the optimal parameters $\alpha$ and $\beta$ for companies and technologies in cybersecurity, depending on the number of companies and linked technologies considered as input.\bigskip
    \end{footnotesize}
\end{table}

\begin{figure}[h!]
    \centering
    \includegraphics[scale=0.22]{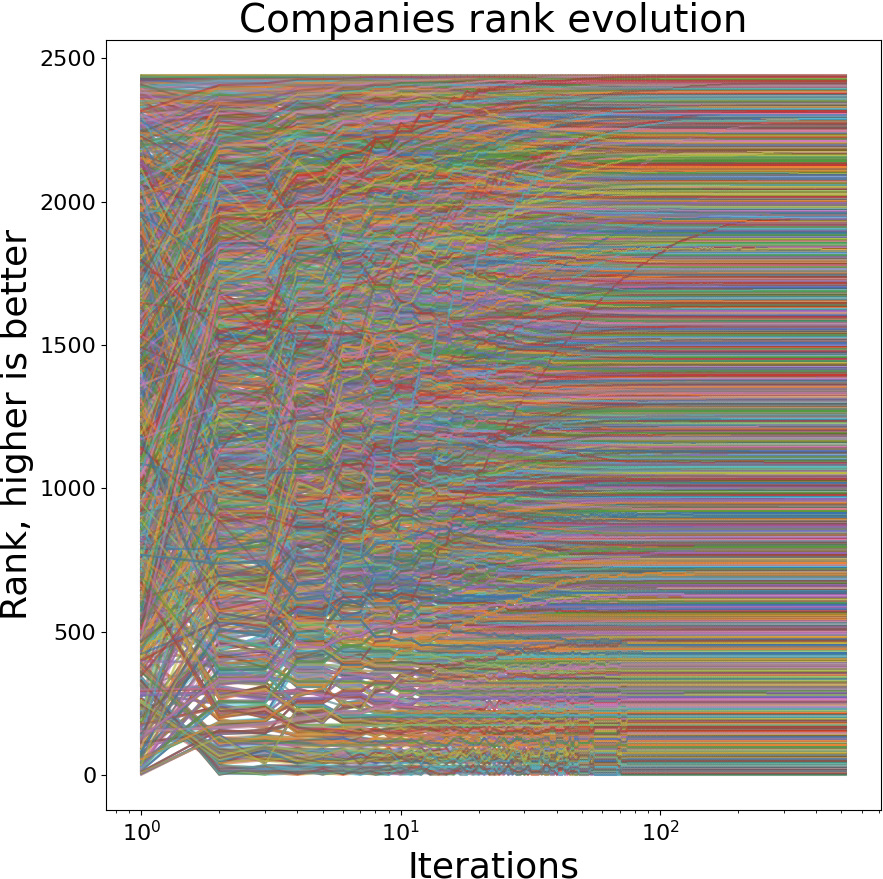}
    \includegraphics[scale=0.22]{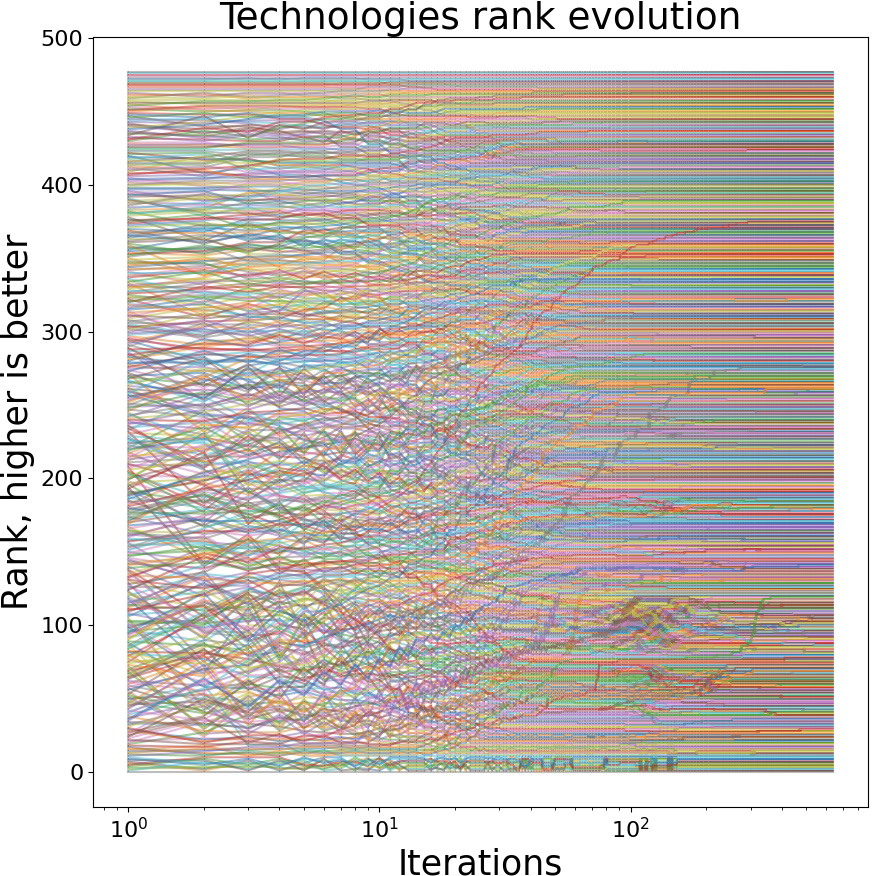}
    \caption{TechRank scores evolution in cybersecurity}
    \bigskip
    \label{fig8}
	\begin{footnotesize}
		This figure displays the TechRank scores evolution over the iterations for 2,429 companies and 477 technologies in cybersecurity.\bigskip
	\end{footnotesize}
\end{figure}

We illustrate the evolution of the TechRank random walker in Figure \ref{fig8}. While the entities' positions significantly change over the first steps, they gradually stabilize. With the 2,429 companies and 477 technologies, the algorithm requires 723 (1,120) iterations for companies (technologies) to converge. Entities starting with a high score (the initialisation is the degree of the node) do not significantly change rank and remain among the best ones. Thus, the algorithm assigns good scores to entities with many neighbours. Instead, entities starting with a low degree may significantly change their score, especially in the case of technologies. TechRank does not only recognize the importance of the most established entities, it also enables to identify emerging technologies. Figure \ref{fig11} shows the first classified entities in cybersecurity.

%
We check how TechRank performs when we change the number of companies and technologies. We fix the number of companies $n^c$, which yields a resulting number of technologies $n^t$. For instance, in the cybersecurity field, by selecting 10 companies randomly, we get 26 technologies. Considering that there are 2,429 cybesecurity-related companies on Crunchbase, we study the runtime running the algorithm for 10, 100, 499, 997, 1494, 1990, and 2,429 companies and 26, 134, 306, 372, 431, 456, and 477 technologies respectively.

\begin{figure}[h!]
    \centering
    \includegraphics[scale=0.32]{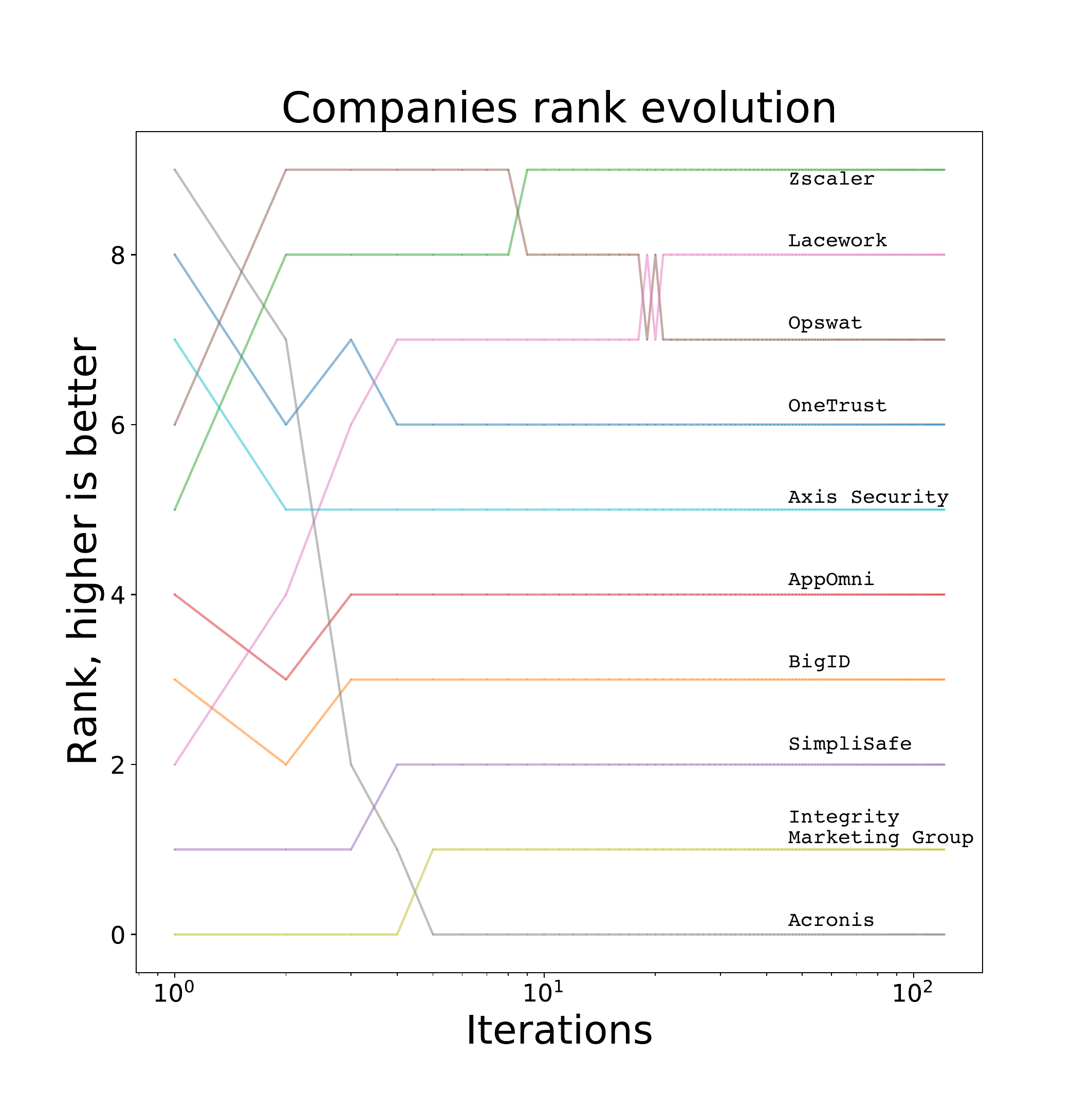}
    \includegraphics[scale=0.32]{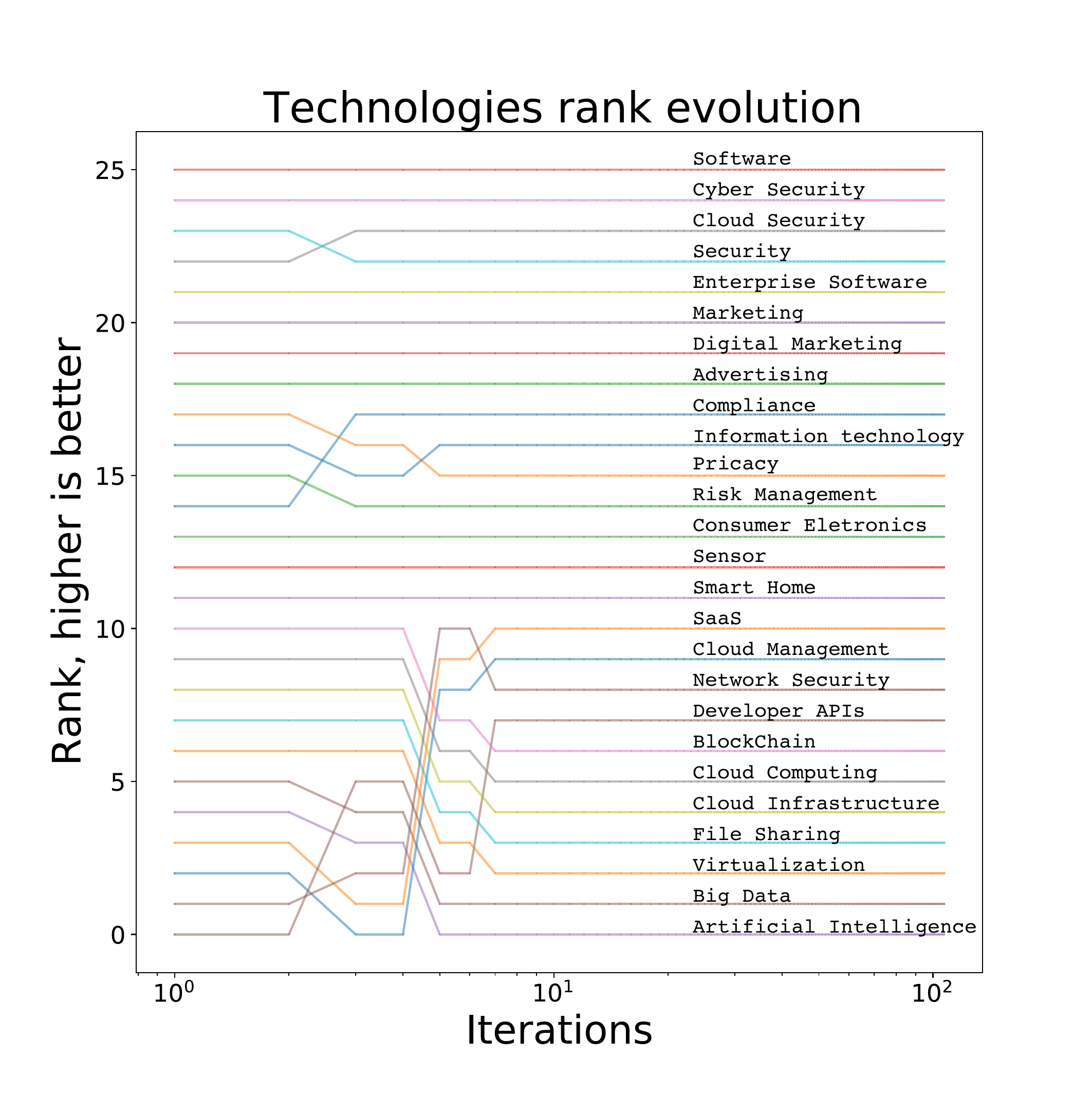}
    \caption{TechRank scores evolution of 10 companies in cybersecurity}
    \bigskip
    \label{fig9}
	\begin{footnotesize}
		This figure displays the TechRank scores evolution over the iterations on a subset of 10 companies and 26 technologies in cybersecurity.\bigskip
	\end{footnotesize}
\end{figure}

\begin{figure}[h!]
    \centering
    \includegraphics[scale=0.30]{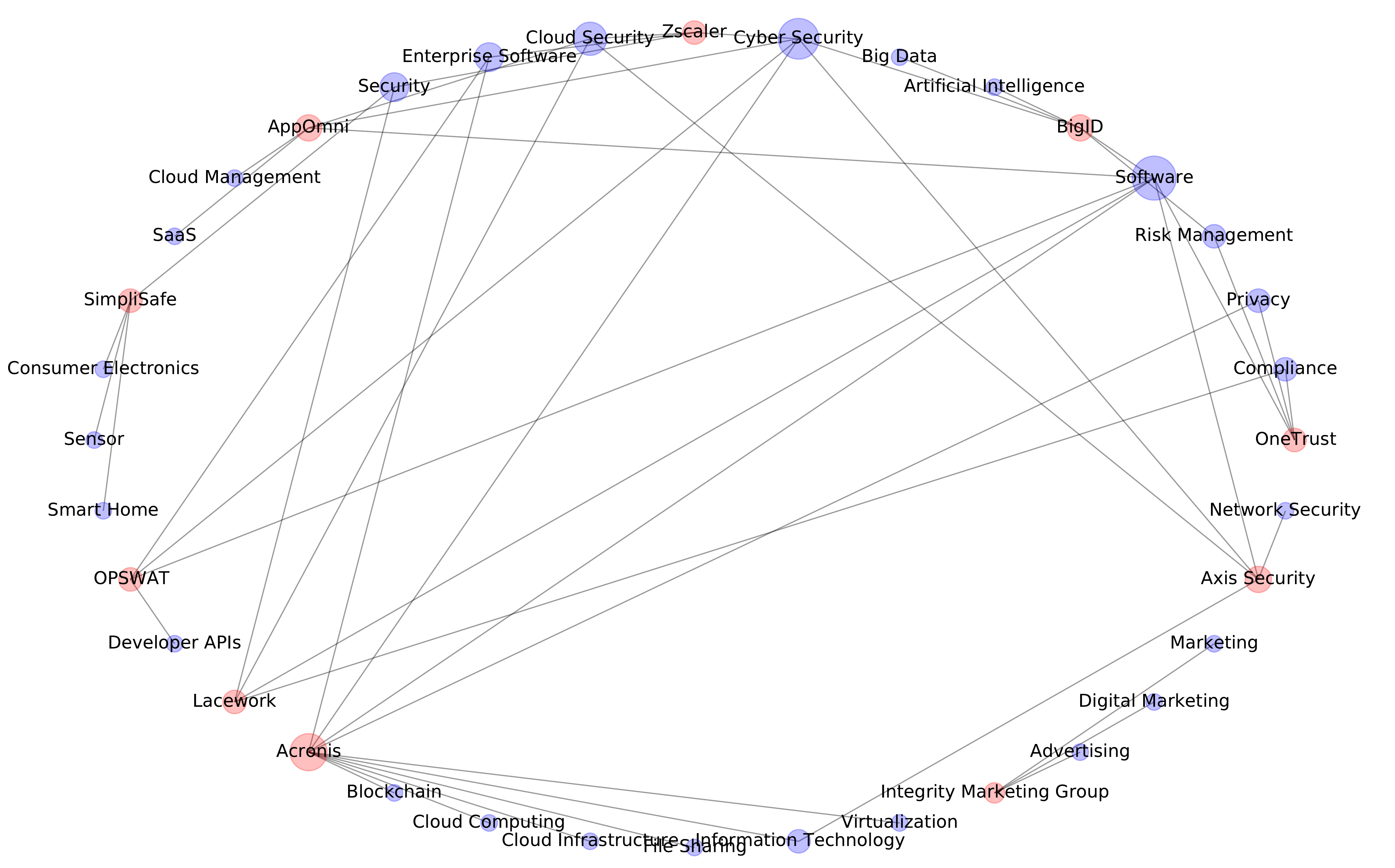}
    \caption{Circular network representation of 10 companies in cybersecurity}
    \bigskip
    \label{fig10}
	\begin{footnotesize}
		This figure displays a circular network representation of a subset of 10 companies and 26 technologies in cybersecurity.\bigskip
	\end{footnotesize}
\end{figure}

\begin{figure}[h!]
    \centering
    \includegraphics[scale=0.3]{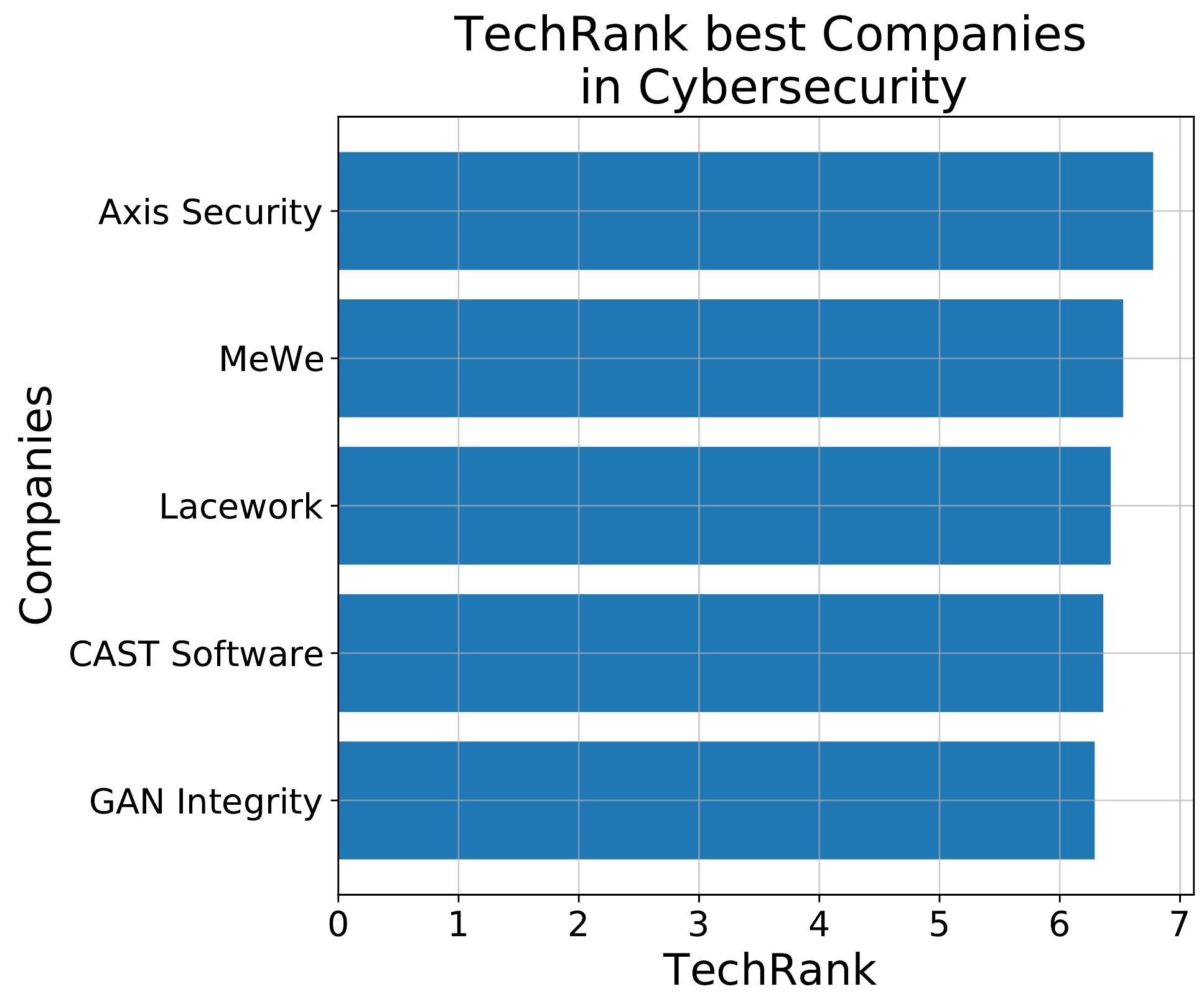}
    \includegraphics[scale=0.3]{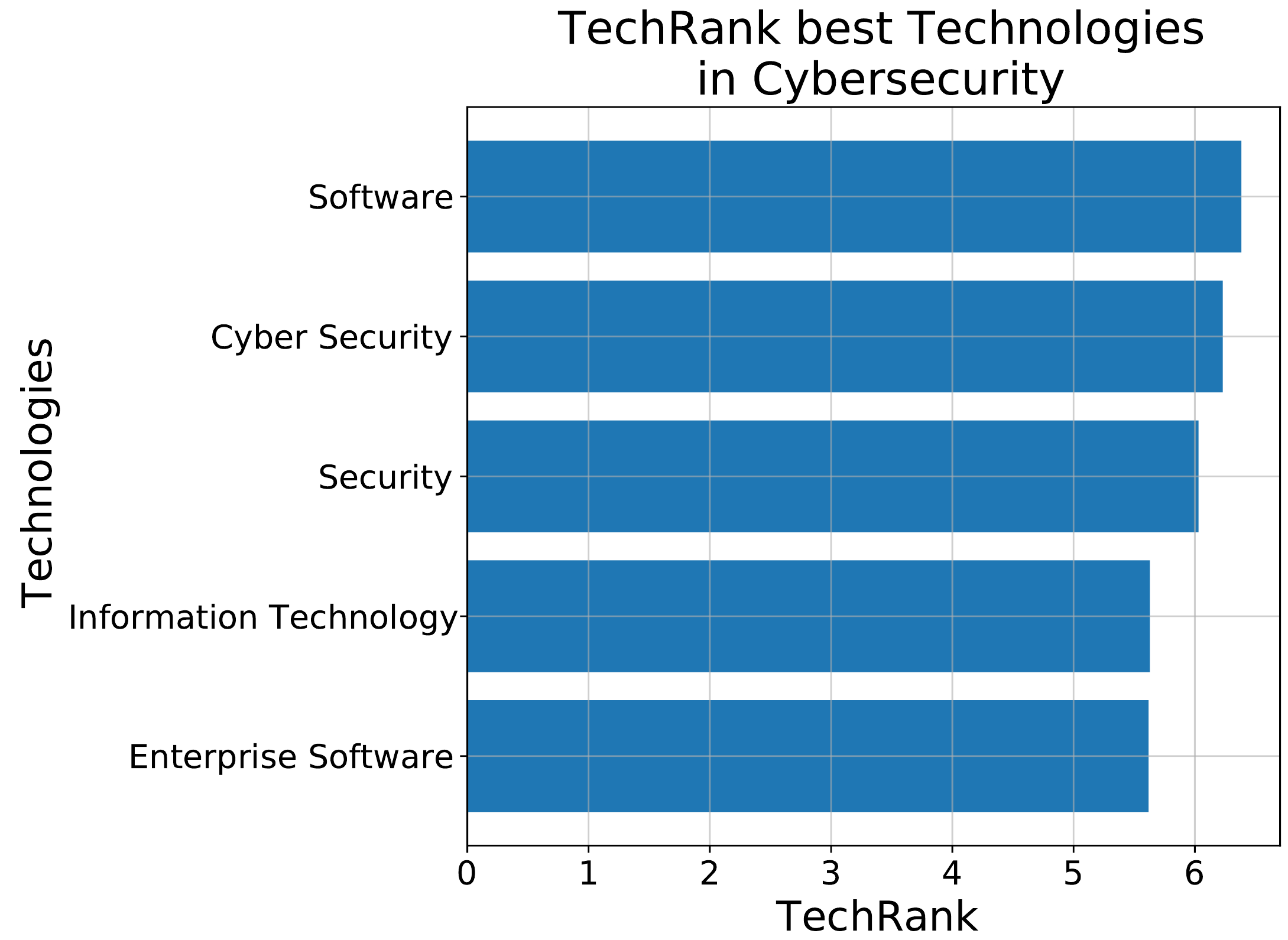}
    \caption{TechRank top five scores in cybersecurity}
    \bigskip
    \label{fig11}
    \begin{footnotesize}
        This left (right) panel displays the top five companies (technologies) according to the TechRank score when run on a subset of 10 companies in cybersecurity.\bigskip
    \end{footnotesize}
\end{figure}

Figure \ref{fig9} displays the results of TechRank applied on a subset of 10 cybersecurity companies. We note that \quotes{AppOmni}'s position does not change over the iteration, while two of its technologies, \quotes{Software as a Service} (Saas) and \quotes{cloud management} increase their scores. In Figure \ref{fig10} we display this restricted network of 10 companies, that shows that SaaS and cloud management do not have other links. Hence, the strength of this company 
depends on its ability to combine important technologies (software, cyber security, and cloud security) with more exotic fields. Similarly, \quotes{Integrity Market Group} is the single involved in some fields (marketing, digital marketing, and advertising). This company does not use more established technologies and thus does not improve its score. Again, in Figure \ref{fig10} we observe that these technologies lie out of the main network. Conversely, \quotes{Lacework} and \quotes{Acronis} follow an opposite trend. Lacework (Acronis) significantly increases (decreases) its score. One explanation for this behaviour is the fact that Acronis is involved in a lot of technologies, most of which are not explored by other companies. On the other hand, Lacework relies on recognized technologies (security, cloud security, and software). Interestingly, the compliance technology, benefits from its connections, increasing its rank by three positions.

\newpage

Table \ref{tab4} reports the number of algorithm iterations before reaching convergence. The number of iterations needed appears to be independent from the number of entities. Technologies need more iterations than companies, which we explain by the fact that there are many more companies than technologies. Since each company has at least one edge, the technology nodes have a higher degree than the companies on average. Thus, we expect the structure and the dynamics related to technologies to be more complex. The algorithm complexity does not only depend on the number of entities, but also on the network structure.

\begin{table}[h!]
    \centering
    \begin{tabular}{cccc}
\multicolumn{2}{c}{Companies} & \multicolumn{2}{c}{Technologies} \\\cmidrule(lr){1-2} \cmidrule(lr){3-4}
Number & Iterations C & Number & Iterations T\\\cmidrule(lr){1-1} \cmidrule(lr){2-2} \cmidrule(lr){3-3} \cmidrule(lr){4-4}
10          & 32                  & 26          & 18                  \\
100         & 100                 & 134         & 155                 \\
499         & 134                 & 306         & 2,469               \\
997         & 196                 & 371         & 194                 \\
1,494       & 180                 & 416         & 871                 \\
1,990       & 240                 & 449         & 5,000               \\
2,429       & 723                 & 477         & 1,120              
    \end{tabular}
    \caption{TechRank convergence}
    \bigskip
    \begin{footnotesize}
		This table reports the number of TechRank iterations before convergence for companies and technologies in cybersecurity.\bigskip
	\end{footnotesize}
    \label{tab4}
\end{table}

\subsection{Investment strategy}
We investigate how investors can select the strategy that reflects their preferences. If investors prefer to focus on technologies, they should choose companies working on the best technologies as selected by the (highest) TechRank score. This decision implies many criteria such as the number of technologies they want to be invested in, the capital allocation for each company, and the diversification. We sketch the procedure to solve this decision process quantitatively in the Appendix \ref{app1}.

\subsection{Comparison with the Crunchbase rank}
Crunchbase assigns a rank to the top companies of each industry, that takes into account the entity's strength of relationships, funding events, news articles, and acquisitions.\footnote{\url{https://about.crunchbase.com/blog/influential-companies/}}  We compare our results in the cybersecurity sector with the Crunchbase rank and investigate the strength of the association between the two scores using Spearman's correlation.

To make the ranks comparable, we convert our algorithm's output into a ranking. The resulting Spearman's correlation of 1.4\% indicates that the two ranks are uncorrelated. We explain these differences by the fact that the Crunchbase rank is fixed, while TechRank is customizable according to investors' preferences. Moreover, the Crunchbase rank focuses on the company's level of activity and not on its market influence. Furthermore, the Crunchbase rank results from an algorithm that involves all the companies, while we focus only on a subset. We attempt to change the investors' preferences and never obtain correlation coefficients above 2\%. Other explanations we propose for this divergence includes the fact we assign a weight which identifies the distance between entities in the ranking. In the same line, TechRank allows decision-makers to set a threshold as starting parameter before running the algorithm. Finally, the Crunchbase algorithm is not open source and we do not know its mechanism, which makes the identification of the divergence's source difficult.

\subsection{Runtime}
We detail the code related to the TechRank algorithm in the Appendix \ref{algo1}. We run it on a machine with a 16-cores Intel Xeon CPU E5-2620 v4 @ 2.10GHz and with 126GB of memory. We investigate the variations in runtime given changes in the number of companies and technologies.

The runtime is a positive function of the number of entities. For technologies the curve is much steeper than that for companies. However, considering that the number of technologies is a direct link  to the number of companies, we repeat the experiment treating companies and technologies together. The random walk phase lines represent the runtime to convergence. There is a strong similarity between the runtime for companies and technologies, which is surprising given their different numbers. This also shows how strongly they are correlated and supports the capability of TechRank to capture the complexity of the cybersecurity technological landscape. Table \ref{tab4} reports all the runtimes and we report the corresponding runtime comparisons for technologies and companies of both cybersecurity and medical field in the Appendix \ref{app3}.\newpage

\begin{table}[h!]
    \centering
    \begin{tabular}{cccccc}
    \multicolumn{3}{c}{Companies} & \multicolumn{3}{c}{Technologies}\\\cmidrule(lr){1-3} \cmidrule(lr){4-6} 
    Number & Parameters' calibration & Convergence & Number & Parameters' calibration & Convergence \\\cmidrule(lr){1-1} \cmidrule(lr){2-2} \cmidrule(lr){3-3} \cmidrule(lr){4-4} \cmidrule(lr){5-5} \cmidrule(lr){6-6}
    10    & 10.21    & 0.56      & 26  & 11.75    & 0.57      \\
    100   & 28.69    & 13.24     & 134 & 35.37    & 13.72     \\
    499   & 189.03   & 470.10    & 306 & 154.79   & 483.25    \\
    997   & 730.43   & 2,023.18  & 371 & 312.65   & 2,392.46  \\
    1,494 & 1,372.17 & 4,514.11  & 416 & 482.18   & 4,404.48  \\
    1,990 & 2,057.42 & 8,396.26  & 449 & 656.95   & 8,096.69  \\
    2,429 & 3,230.99 & 16,890.26 & 477 & 1,071.84 & 12,779.62
    \end{tabular}
    \caption{TechRank runtime}
    \bigskip
    \label{tab5}
    \begin{footnotesize}
		This table reports the TechRank runtime for companies and technologies in cybersecurity.\bigskip
	\end{footnotesize}
\end{table}

\subsection{Exogenous factors}
We conduct a sensitivity analysis based on investors' preferences. We restrict the analysis to 1,000 companies only, given the to long runtime required. We assume investors to be interested in firm location only and consider the case of an investor based in New York City and in San Francisco, in turn. In Table \ref{tab6} we report the outcome in terms of location for the five top ranked companies in both cases. We uncover a location change in the company ranking, with the first one being in the state of New York (investor based in New York City) and in the state of California (investor based in San Francisco), respectively. The companies with lower rank also reflects these geographical preferences, albeit with some exceptions (Singapore and Beijing). This implies that, even if remote companies are disadvantaged, their other attributes overcome this flaw.

\begin{table}[h!]
    \begin{tabular}{lccccc}
        \makecell{Company rank/\\ Investor location} & 1 & 2 & 3 & 4 & 5 \\ \cmidrule(lr){1-1} \cmidrule(lr){2-2} \cmidrule(lr){3-3} \cmidrule(lr){4-4} \cmidrule(lr){5-5} \cmidrule(lr){6-6}
        New York City & \makecell{New York City\\ (USA)} & \makecell{Massachusetts\\ (USA)} & \makecell{Quebec\\ (Canada)} & \makecell{California\\ (USA)} & \makecell{Singapore\\ (Singapore)}\\
        San Francisco & \makecell{California\\  (USA)} & \makecell{Illinois\\  (USA)} & \makecell{California\\  (USA)} & \makecell{Beijing\\ (China)} & \makecell{Arizona\\ (USA)} \\
    \end{tabular}
    \centering
    \caption{Companies TechRank scores with location}
    \bigskip
    \label{tab6}
     \begin{footnotesize}
		This table reports the location of the top five TechRank companies' scores when the geographical preference is fully set in the location of the investors (New York City and San Francisco).\bigskip
	\end{footnotesize}
\end{table}
\subsection{Robustness tests}

\begin{figure}[h!]
    \centering
    \includegraphics[scale=0.2]{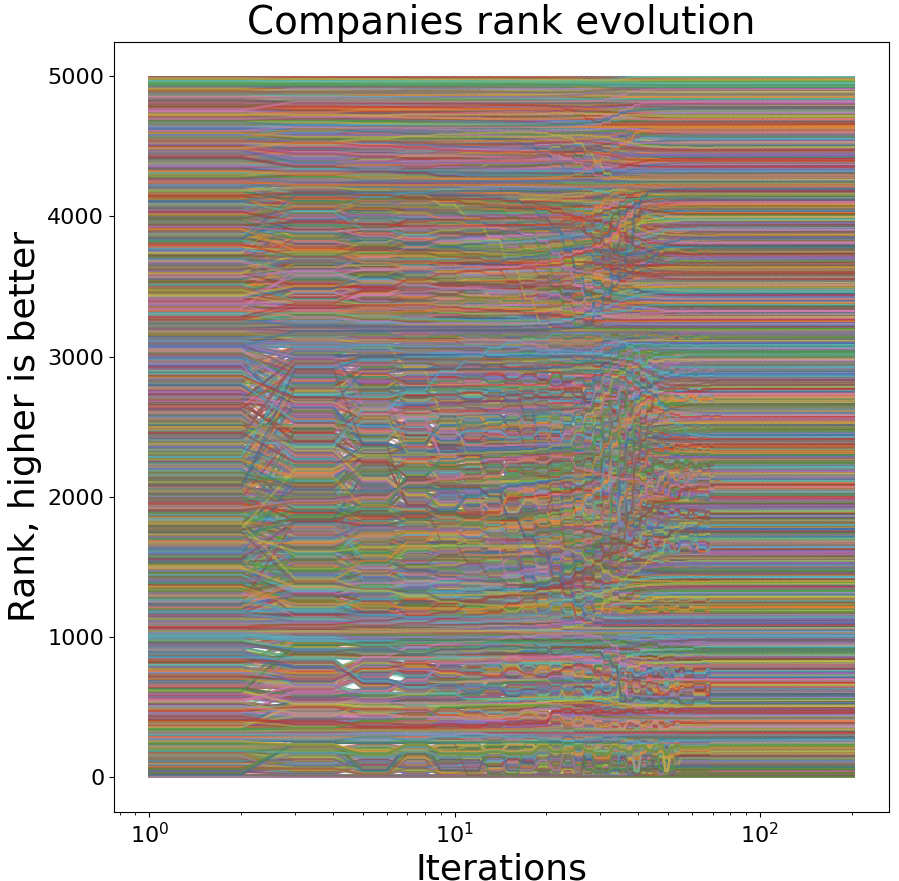}
    \includegraphics[scale=0.2]{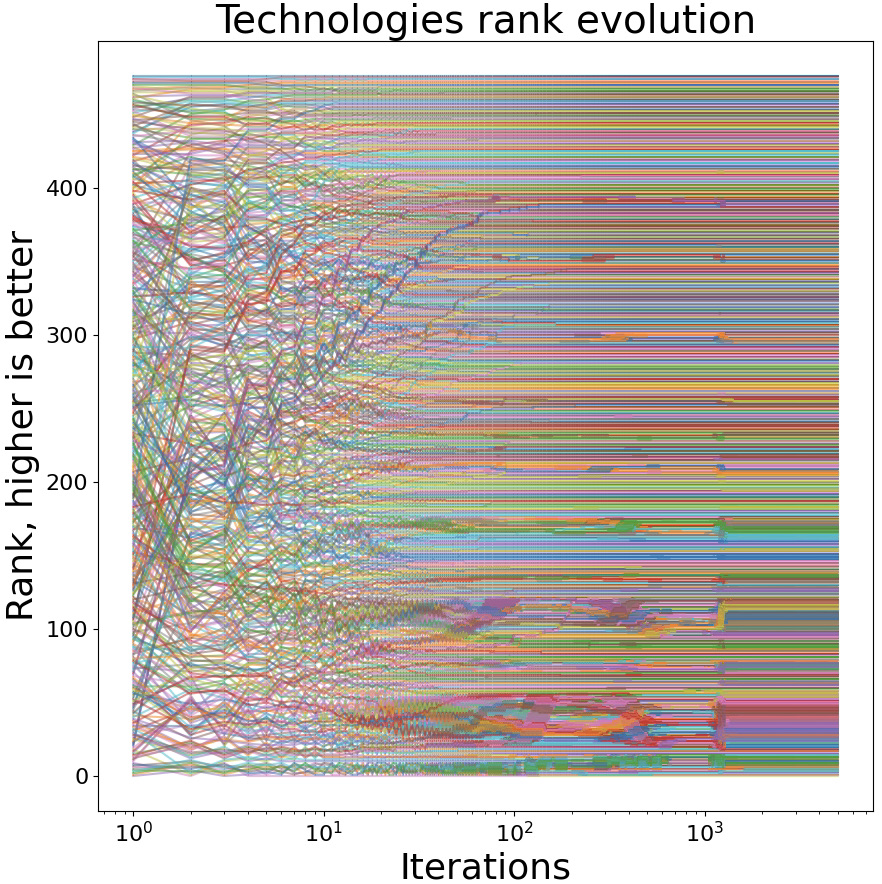}
    \caption{TechRank scores evolution in the medical field}
    \bigskip
    \label{fig12}
    \begin{footnotesize}
        This figure displays the TechRank scores evolution over the iterations for 2,429 companies and 477 technologies in the medical field.\bigskip
    \end{footnotesize}
\end{figure}

To test the robustness of our algorithm and benchmark the cybersecurity sector, we apply TechRank for companies in the medical sector. We choose this sector given the important number of companies (twice the number of companies working in cybersecurity). We select the companies with the same methodology, which yields a total of 4,996 companies and 437 technologies. Figure \ref{fig12} shows the results of TechRank in the medical sector. The runtime for these companies, reported in the Appendix \ref{app3}, is on par with those of the cybersecurity sector. To make the two fields comparable, we set as x-label the number of entities, for both companies and technologies. The results reveal that the runtime of the two fields, for both the parameter calibration and the random walker steps, follow the same behaviour, for both companies and technologies. Increasing the number of entities does not yield any significant change both in terms of convergence and runtime behaviour. Finally, unlike Klein et al. (2015), for which the $\alpha$ remains constant and $\beta$ changes significantly, we observe that all of our parameters significantly change across sectors\cite{KleinMaillartChuang2015}.

\section{Conclusion}\label{sec5}

\subsection{Limitations}\label{limitations}
We choose technologies related to cybersecurity according to Crunchbase description, \textit{e.g.}, security, privacy, or confidentiality. Since these words may overlap other fields, we require the description to contain at least two of them to classify a company as cybersecurity. This naive strategy could be improved with more sophisticated techniques, such as natural language processing NLP. We also face limitation given the lack of information about companies' resources allocation towards technologies. We only have a list of technologies for each company without more information. Instead, it would be helpful to observe the amount of expenditures towards each technologies.
Our algorithm is static as we do not have access to time series and it would be interesting to study how the bi-partite network changes. Finally, we are well aware that introducing exogenous variables may induce a bias given potential outliers. Our normalization procedure divides the factors by their maximum, which may lead to unproportionate results if the maximum is an outlier. However, we do not believe that removing outliers is a viable solution, since this would lead to overlook potentially profitable opportunities.

\subsection{Further research}\label{further_research}

This research can be expanded with time series, to investigate, for instance, the outcome of a company divesting from a technology, or investing in a new one. This would also enables to assess which new technologies are successful. This would give more insights about investment choices towards too recent ideas. A focus on percolation theory could help assess the effects of a node disappearance in the network \cite{PiraveenanProkopenkoHossain2013}. For this purpose, machine learning methods could also be employed. Further research should be devoted to investigate additional exogenous factors in TechRank, enabling investors the widest range of features available possible. Alternative exogenous factors include the inception date of the company, number of employees, social networks activity, and even the Crunchbase rank, which is itself based upon entity's strength of relations, funding events, news articles, or acquisitions. By the same token, further investigation should confirm the pertinence of the TechRank algorithm in fields that include more entities, as an increase in nodes could lead to coordination problems. Finally, it would be crucial to assess the long-term effects of the TechRank algorithm on investments returns and technologies' development through back-testing, that once again, requires time series.

\subsection{Conclusion}
We introduce TechRank, an algorithm that assigns a score to companies and technologies in complex systems. This methodology constitutes the first step towards a new data-driven investment strategy, which enables investors to follow their preferences while benefiting from a quantitative approach. We include investors' preference based upon a case-by-case study. Our algorithm convergence depends on the number of entities and the complexity of the relationships within the bi-partite network. Using a restricted number of companies in cybersecurity, we analyze the TechRank scores and we explain the score variations of entities over iterations. We also explore how results change depending on the company's location. Finally, we conduct robustness tests in the medical field, for which our results are qualitatively similar. 

We believe that our approach brings value to help investors' form their decisions. Moreover, our algorithm's flexibility allows to include exogenous factors and preferences, which is impossible in alternative existing company ranks, such as that of Crunchbase. Given our algorithm performance for cybersecurity, a highly complex market, as a case study, we believe that our algorithm would perform well in all markets. TechRank is a complementary, if not alternative, way to look at portfolio choices.


\clearpage
\bibliographystyle{styles/jfe}
\bibliography{references.bib}

\begin{thebibliography}{33}
\expandafter\ifx\csname natexlab\endcsname\relax\def\natexlab#1{#1}\fi

\bibitem[{Battiston et~al.(2014)Battiston, Nicosia, and
  Latora}]{BattistonNicosiaLatora2014}
Battiston, F., Nicosia, V., Latora, V., 2014. Structural measures for multiplex
  networks. Physical Review E 89, 1--14.

\bibitem[{Bavelas(1948)}]{Bavelas1948}
Bavelas, A., 1948. A mathematical model for group structures. Applied
  Anthropology 7 (3), 16--30.

\bibitem[{Benzi et~al.(2013)Benzi, Estrada, and
  Klymko}]{BenziEstradaKlymko2013}
Benzi, M., Estrada, E., Klymko, C., 2013. Ranking hubs and authorities using
  matrix functions. Linear Algebra and its Applications 438, 2447--2474.

\bibitem[{Besten~\(den\)(2021)}]{Besten2021}
Besten~\(den\), M.~L., 2021. Crunchbase research: Monitoring entrepreneurship
  research in the age of big data. Available at
  \url{http://dx.doi.org/10.2139/ssrn.3724395}\EatDot .

\bibitem[{Bonacich(1972)}]{Bonacich1972}
Bonacich, P., 1972. Factoring and weighting approaches to status scores and
  clique identification. Journal of Mathematical Sociology 2, 113--120.

\bibitem[{Buldyrev et~al.(2010)Buldyrev, Parshani, Paul, Stanley, and
  Shlomo}]{BuldyrevParshaniPaulStanleyShlomo2010}
Buldyrev, S.~V., Parshani, R., Paul, G., Stanley, H.~E., Shlomo, H., 2010.
  Catastrophic cascade of failures in interdependent networks. Nature 464,
  1025–--1028.

\bibitem[{Canito et~al.(2018)Canito, Ramos, Moro, and
  Rita}]{CanitoRamosMoroRita2018}
Canito, J., Ramos, P., Moro, S., Rita, P., 2018. Unfolding the relations
  between companies and technologies under the {B}ig {D}ata umbrella. Computers
  in Industry 99, 1--8.

\bibitem[{Christensen(1997)}]{Christensen1997}
Christensen, C.~M., 1997. The Innovators Dilemma: When New Technologies Cause
  Great Firms to Fail. Harvard Business School Press, Boston, MA.

\bibitem[{Cochrane(2005)}]{Cochrane2005}
Cochrane, J.~H., 2005. The risk and return of venture capital. Journal of
  Financial Economics 75, 3--52.

\bibitem[{Dalle et~al.(2017{\natexlab{a}})Dalle, Besten~\(den\), and
  Menon}]{DalleBestenMenon2017W}
Dalle, J.-M., Besten~\(den\), M.~L., Menon, C., 2017{\natexlab{a}}. Using
  {C}runchbase for economic and managerial research. Available at
  \url{https://doi.org/10.1787/18151965}\EatDot .

\bibitem[{Dalle et~al.(2017{\natexlab{b}})Dalle, Besten~\(den\), and
  Menon}]{DalleBestenMenon2017}
Dalle, J.-M., Besten~\(den\), M.~L., Menon, C., 2017{\natexlab{b}}. Using
  {C}runchbase for economic and managerial research. Available at
  \url{https://doi.org/10.1787/18151965}\EatDot .

\bibitem[{Donato et~al.(2004)Donato, Laura, Leonardi, and
  Millozzi}]{DonatoLauraLeonardiMillozzi2004}
Donato, D., Laura, L., Leonardi, S., Millozzi, S., 2004. Large scale properties
  of the {W}ebgraph. European Physical Journal B 38, 239--243.

\bibitem[{Ewens(2009)}]{Ewens2009}
Ewens, M., 2009. A new model of venture capital risk and return. Available at
  \url{http://dx.doi.org/10.2139/ssrn.1356322}\EatDot .

\bibitem[{Freeman(1978)}]{Freeman1978}
Freeman, L.~C., 1978. Centrality in social networks conceptual clarification.
  Social Networks 1, 215--239.

\bibitem[{Gold et~al.(2001)Gold, Malhotra, and Segars}]{GoldMalhotraSegars2001}
Gold, A.~H., Malhotra, A., Segars, A.~H., 2001. Knowledge management: An
  organizational capabilities perspective. Journal of Management Information
  Systems 18, 185--214.

\bibitem[{Gordon et~al.(2018)Gordon, Loeb, Lucyshyn, and
  Zhou}]{GordonLoebLucyshynZhou2018}
Gordon, L.~A., Loeb, M.~P., Lucyshyn, W., Zhou, L., 2018. Empirical evidence on
  the determinants of cybersecurity investments in private sector firms.
  Journal of Information Security 9, 133--153.

\bibitem[{Gornall and Strebulaev(2020)}]{GornallStrebulaev2020}
Gornall, W., Strebulaev, I.~A., 2020. Squaring venture capital valuations with
  reality. Journal of Financial Economics 135, 120--143.

\bibitem[{Hidalgo et~al.(2009)Hidalgo, Hausmann, and
  Dasgupta}]{HidalgoHausmannDasgupta2009}
Hidalgo, C.~A., Hausmann, R., Dasgupta, P.~S., 2009. The building blocks of
  economic complexity. Proceedings of the National Academy of Sciences 26,
  10570--10575.

\bibitem[{Ingole and Nichat(2013)}]{IngoleNichat2013}
Ingole, P.~V., Nichat, M.~K., 2013. Landmark based shortest path detection by
  using dijkestra algorithm and haversine formula. International Journal of
  Engineering Research and Applications (IJERA) 3, 162--165.

\bibitem[{Katz(1953)}]{Katz1953}
Katz, L., 1953. A new status index derived from sociometric analysis.
  Psychometrika 18, 39--43.

\bibitem[{Klein et~al.(2015)Klein, Maillart, and
  Chuang}]{KleinMaillartChuang2015}
Klein, M., Maillart, T., Chuang, J., 2015. The virtuous circle of {W}ikipedia:
  Recursive measures of collaboration structures. In: {\em Proceedings of the
  18th ACM Conference on Computer Supported Cooperative Work \& Social
  Computing\/},  1106--1115.

\bibitem[{Korteweg and Nagel(2016)}]{KortewegNagel2016}
Korteweg, A., Nagel, S., 2016. Risk‐adjusting the returns to venture capital.
  Journal of Finance 71, 1437--1470.

\bibitem[{Kurant and Thiran(2006)}]{KurantThiran2006}
Kurant, M., Thiran, P., 2006. Layered complex networks. Physical Review Letters
  96 (13), 1--4.

\bibitem[{Moskowitz and
  Vissing-Jørgensen(2002)}]{MoskowitzVissing-Jorgensen2002}
Moskowitz, T.~J., Vissing-Jørgensen, A., 2002. The returns to entrepreneurial
  investment: A private equity premium puzzle? American Economic Review 92,
  745--778.

\bibitem[{Page et~al.(1999)Page, Brin, Motwani, and
  Winograd}]{PageBrinMotwaniWinograd1999}
Page, L., Brin, S., Motwani, R., Winograd, T., 1999. The {P}age{R}ank citation
  ranking: Bringing order to the web. Available at
  \url{http://ilpubs.stanford.edu:8090/422/}\EatDot .

\bibitem[{Peng(2001)}]{Peng2001W}
Peng, L., 2001. Building a venture capital index. Available at
  \url{http://dx.doi.org/10.2139/ssrn.281804}\EatDot .

\bibitem[{Piraveenan et~al.(2013)Piraveenan, Prokopenko, and
  Hossain}]{PiraveenanProkopenkoHossain2013}
Piraveenan, M., Prokopenko, M., Hossain, L., 2013. Percolation centrality:
  Quantifying graph-theoretic impact of nodes during percolation in networks.
  PLOS ONE 8 (1), 1--14.

\bibitem[{Saxena and Iyengar(2020)}]{SaxenaIyengar2020W}
Saxena, A., Iyengar, S., 2020. Centrality measures in complex networks: A
  survey. Available at \url{https://doi.org/10.48550/arXiv.2011.07190}\EatDot .

\bibitem[{Saxena and Iyengar(2017)}]{SaxenaIyengar2017W}
Saxena, A., Iyengar, S. R.~S., 2017. Global rank estimation. Available at
  \url{https://arxiv.org/abs/1710.11341}\EatDot .

\bibitem[{Schwarty and Moon(2000)}]{SchwartzMoon2000}
Schwarty, E.~S., Moon, M., 2000. Rational pricing of internet companies.
  Financial Analysts Journal 56 (3), 62--75.

\bibitem[{Tu et~al.(2018)Tu, Jiang, Song, and Zhang}]{TuJiangSongZhang2018}
Tu, X., Jiang, G.-P., Song, Y., Zhang, X., 2018. Novel multiplex {P}age{R}ank
  in multilayer networks. IEEE Access 6, 12530--12538.

\bibitem[{Xing and Ghorbani(2004)}]{XingGhorbani2004}
Xing, W., Ghorbani, A., 2004. Weighted {P}age{R}ank algorithm. In: {\em Second
  Annual Conference on Communication Networks and Services Research\/},
  305--314.

\bibitem[{Zhong et~al.(2018)Zhong, Chuanren, Zhong, and
  Xiong}]{ZhongChuanrenZhongXiong2018}
Zhong, H., Chuanren, L., Zhong, J., Xiong, H., 2018. Which startup to invest
  in: A personalized portfolio strategy. Annals of Operations Research 263,
  339--360.

\end{thebibliography}
\clearpage
\newpage

\section*{Appendix}\label{sec6}

\setcounter{table}{0}
\renewcommand{\thetable}{A\arabic{table}}
\setcounter{figure}{0}
\renewcommand{\thefigure}{A\arabic{figure}}

\begin{figure}[h!]
    \centering
    \includegraphics[width=0.5\textwidth]{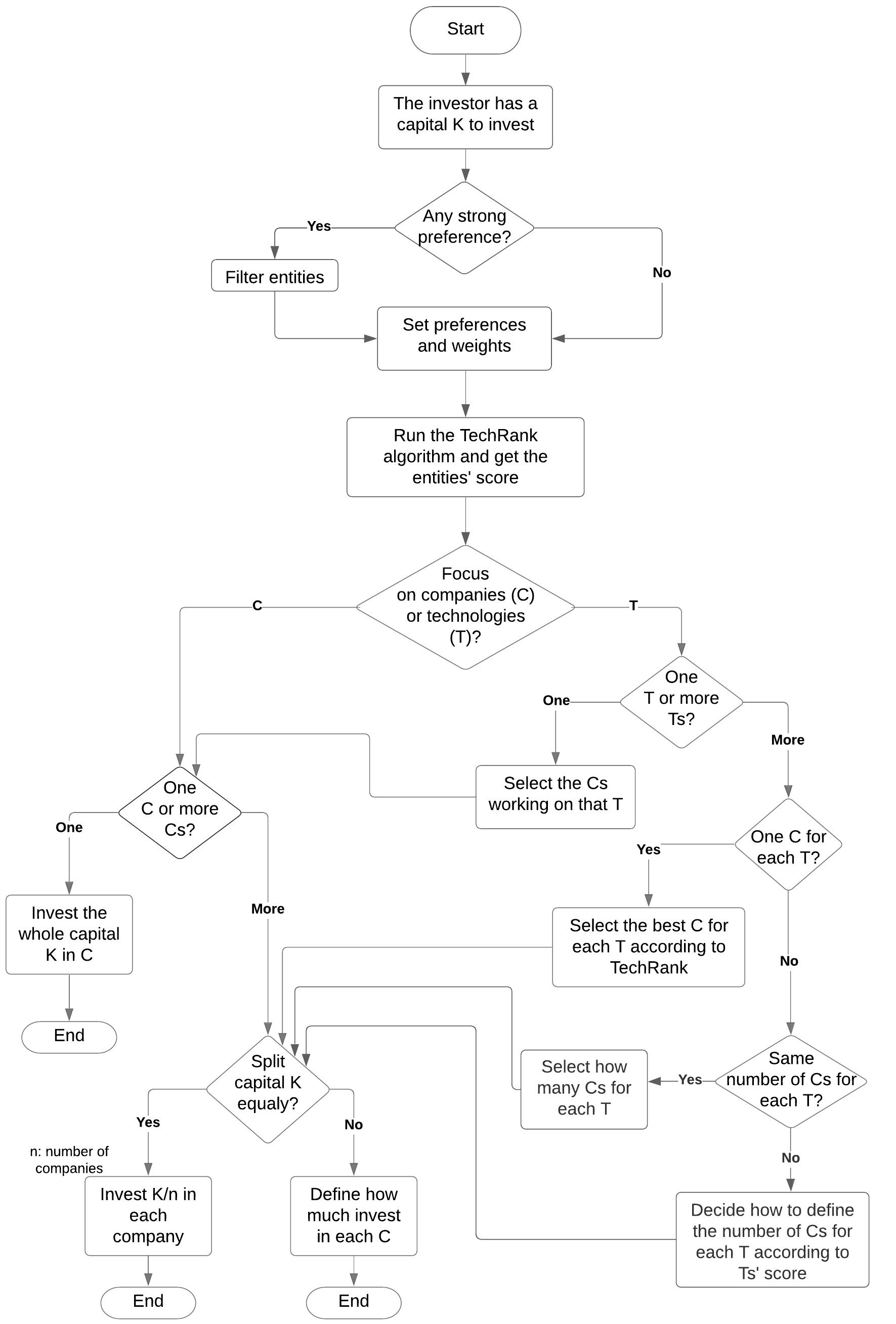}
    \caption{Flowchart of the investment process}
    \bigskip
    \label{app1}
    \begin{footnotesize}
        This flowchart sketches a potential investment process that uses TeckRank and investment preferences (exogenous factors and investment styles) before reaching an optimal investment and portfolio choice.\bigskip
    \end{footnotesize}
\end{figure}

\subsection*{List of words used to identify companies' sectors}
\label{app2}
List of words related to cybersecurity: \textit{cybersecurity, confidentiality, integrity, availability, secure, security, safe, reliability, dependability, confidential, confidentiality, integrity, availability, defence, defensive, privacy.}\bigskip

List of words related to the medical field: \textit{cure, medicine, surgery, doctors, nurses, hospital, medication, prescription, pill, health, cancer, antibiotic, HIV, cancers, disease, resonance, rays, CAT, blood, blood transfusion, accident, injuries, emergency, poison, transplant, biotechnology, health care, healthcare, health-tech, genetics, DNA, RNA, lab, heart, lung, lungs, kidneys, brain, gynaecologist, cholesterol, diabetes, stroke, infections, infection, ECG, sonogram}.

\begin{figure}[h!]
    \centering
    \includegraphics[scale=0.3]{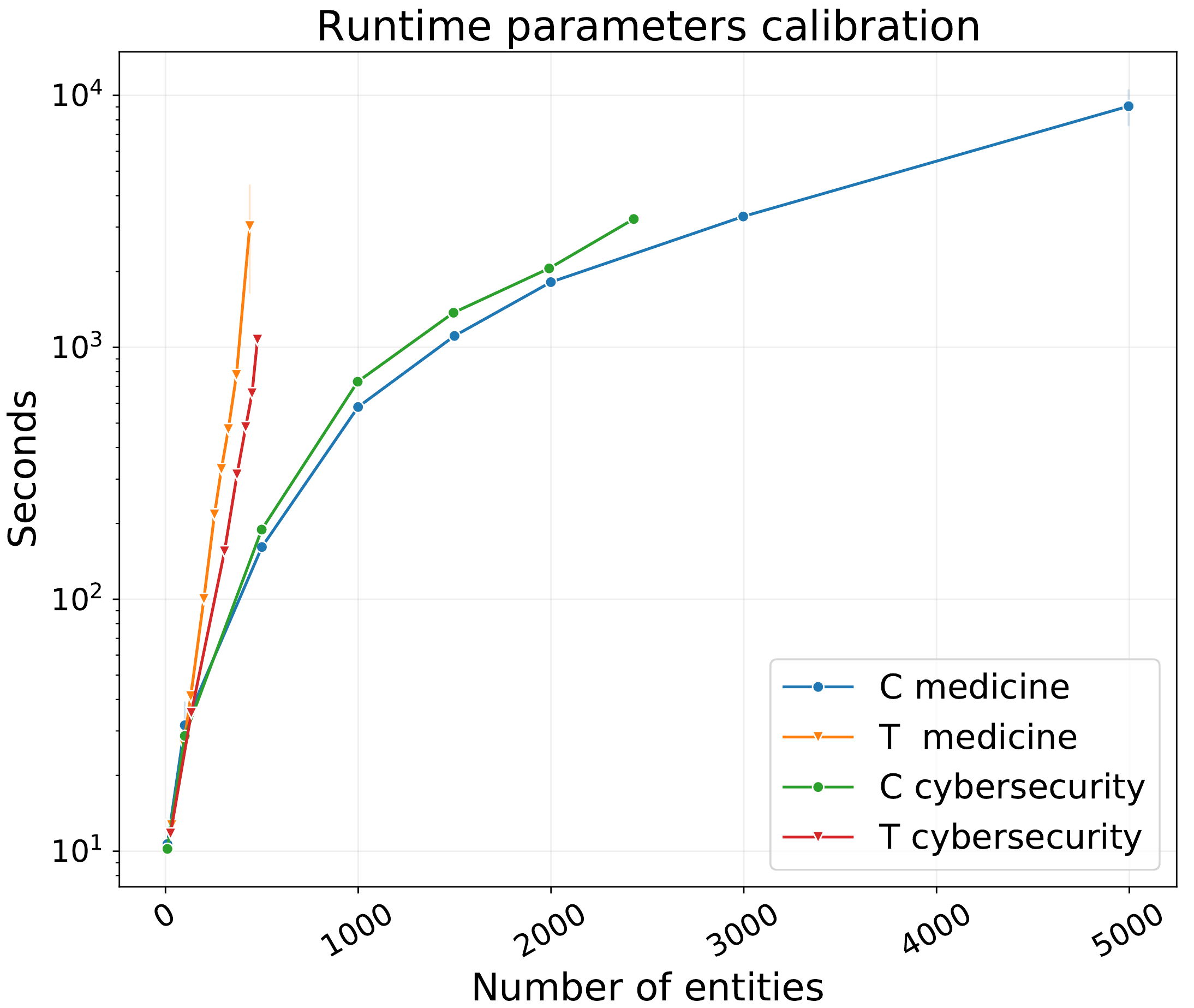}
    \includegraphics[scale=0.3]{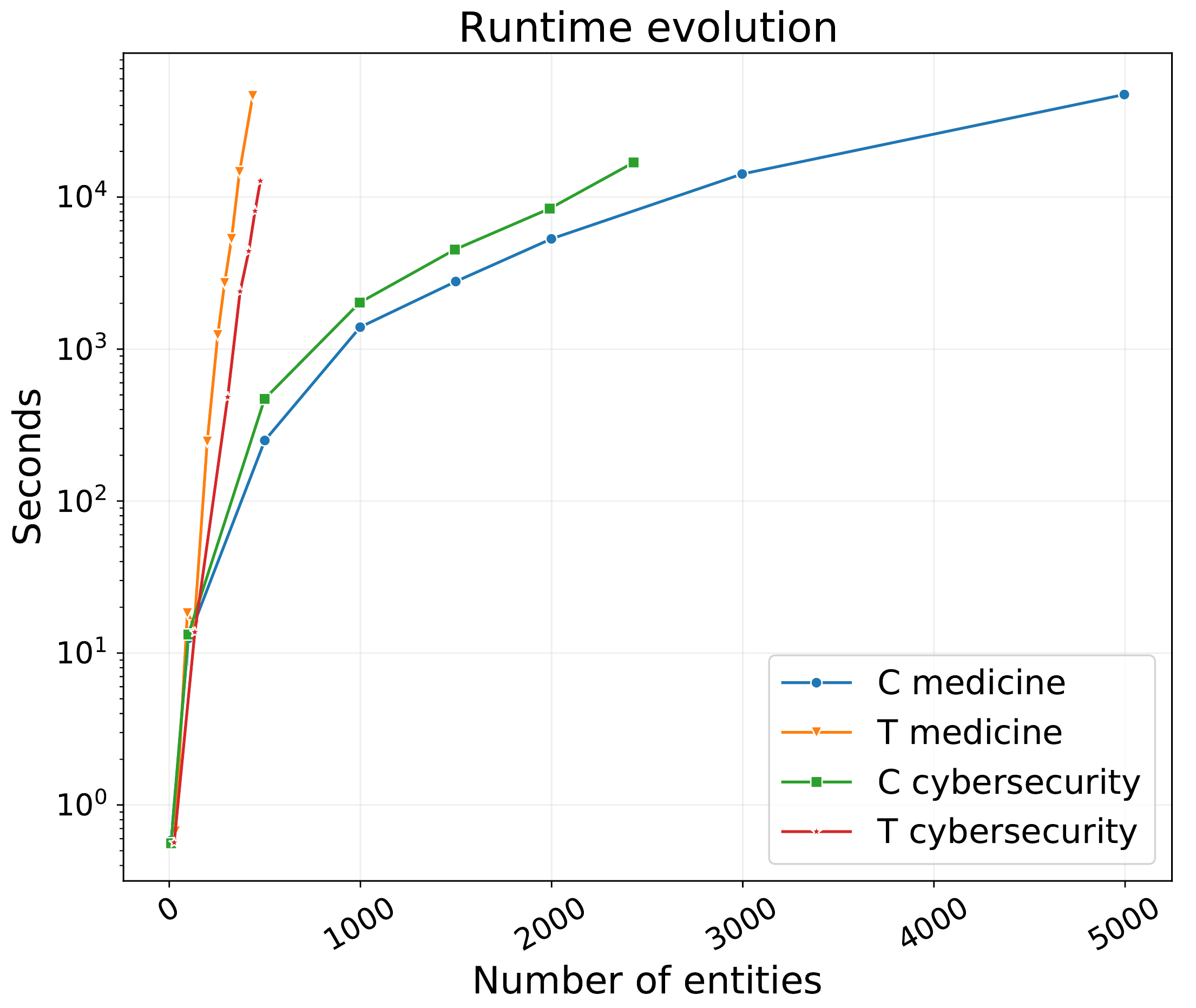}
    \caption{Runtime comparisons}
    \bigskip
    \label{app3}
    \begin{footnotesize}
        The left panel displays the grid search runtime for cybersecurity and medical fields. The right panel displays the parameters' calibration runtime for the cybersecurity and medical fields. The ordinate axis uses a logarithmic scale.\bigskip
    \end{footnotesize}
\end{figure}
\newpage

\begin{algorithm}[h!]
  \begin{algorithmic}[1]
      \State $e^C \gets [0] \cdot len(c\_names)$
      \For{$c \in range(c\_names)$}
          \For{$i \in range(i\_names)$} 
              \For{$c \in range(i\_names)$} 
                \State $e^{IC}_{i,c} \gets \sum_{t=0}^{\mathcal{T}} \gamma^{i,c}_t $ \Comment{$\gamma^{t}_{i,c}$ is the amount of the investment from $i$ to $c$ at time $t$}
                \State $ e^C[c] \gets e^C[c] + e^{IC}_{i,c} $
              \EndFor
          \EndFor
      \EndFor
      \State $e^C_{max} \gets \max{(e^C)}$
      \State $f^C \gets e^C/e_{max}$ \Comment{$f^C$: list of previous investments for each technology}
      \State \textbf{return} $f^C$
  \end{algorithmic}
   \caption{Previous investments factor for companies}
  \label{algo1}
\end{algorithm}

\begin{algorithm}[h!]
  \begin{algorithmic}[1]
      \State $e^C \gets [0] \cdot len(c\_names)$
      \For{$c \in range(c\_names)$}
          \For{$i \in range(i\_names)$} 
            \State $e^{IC}_{i,c} \gets \sum_{t=0}^{(T)} \gamma^{IC}_t $ \Comment{$\gamma^{t}_{i,c}$ is the amount of the investment from $i$ to $c$ at time $t$}
            \State $ e^C[c] \gets e^C[c] + e^{IC}_{i,c} $
          \EndFor
      \EndFor
      \State $e^T \gets e^C \cdot M^{CT}$ \Comment{Matrix multiplication}
      \State $e_{max} \gets \max{(e^T)}$
      \State $f^T \gets e^T/e_{max}$ \Comment{$f^T$: list of previous investments for each technology}
      \State \textbf{return} $f^T$
  \end{algorithmic}
  \caption{Previous investments factor for technologies}
  \label{algo2}
\end{algorithm}

\begin{algorithm}[h!]
  \begin{algorithmic}[1]
      \State $h\_dict \gets \{\}$
      \For{$c\_name, c\_address \in c\_locations$}
        \State $lat \gets c\_address.latitude$
        \State $lon \gets c\_address.longitude$
        \State $h \gets haver\_dist(lat, lon, lat\_inv, lon\_in)$ \Comment{haver\_dist is a function we have created}
        \State $h\_dict[c\_name] \gets 1/h$
      \EndFor
      \State $h\_max \gets \max{(h\_dict)}$
      \For{$c\_name, h \in h\_dict$}
        \State $h\_dict[c\_name] \gets 1 - h/h\_max$
      \EndFor
      \State \textbf{return} $h\_dict$
  \end{algorithmic}
   \caption{Geographic coordinates factor}
  \label{algo3}
\end{algorithm}

\newpage
\subsection*{Distance computation}
\label{app4}
We obtain the distance between two points on earth with the Haversine approximation ($\hav(\theta)$), using latitude and longitude of the locations\cite{IngoleNichat2013}.

Let $(\lambda_1, \phi_1)$ and $(\lambda_2, \phi_2)$ be the longitude and latitude in radiance of two points on a sphere and $\theta$ the central angle given by the spherical law of cosines, the Haversine distance writes,

\begin{equation}\label{eq11}
    h = \hav(\theta) = \hav(\phi_2- \phi_1) + \cos{\phi_1}\cos{\phi_2} \hav(\lambda_2-\lambda_1).
\end{equation}

\end{document}